\documentclass[twocolumn,numberedappendix,twocolappendix,appendixfloats]{openjournal_dhayaa}

\usepackage{amsmath}
\usepackage{fontawesome5}
\usepackage{color}
\usepackage{xcolor}

\usepackage{natbib}
\setcitestyle{aysep={}}

\usepackage[colorlinks=true
  ,urlcolor=blue
  ,anchorcolor=blue
  ,citecolor=blue
  ,filecolor=blue
  ,linkcolor=blue
  ,menucolor=blue
  ,linktocpage=true
  ,pdfproducer=medialab
  ,pdfa=true
]{hyperref}

\usepackage{graphicx}	
\usepackage{amsmath}	

\usepackage{fontawesome5}
\usepackage{color}
\usepackage{xcolor}
\usepackage{xspace}

\newcommand{\eg}{{\sl e.g.}, }   
\newcommand{\ie}{{\sl i.e.}, }

\newcommand{\decade}{\textsc{DECADE}\xspace}
\newcommand{\cosmocov}{\textsc{CosmoCov}\xspace}

\definecolor{orcidlogocol}{HTML}{A6CE39}
\definecolor{purple}{RGB}{128, 0, 128}

\newcommand{\OrcidID}[1]{ \href[urlcolor = red]{https://orcid.org/#1}{\textcolor{lightgray}{\faOrcid}}}
\newcommand{\OrcidIDName}[2]{\href{https://orcid.org/#1}{#2}}

\newcommand{\Om}{\Omega_{\rm m}}
\newcommand{\Seight}{S_8}

\defcitealias{y3-shapecatalog}{\textsc{GS21}}
\defcitealias{paper1}{\textsc{Paper~I}}
\defcitealias{paper2}{\textsc{Paper~II}}
\defcitealias{paper3}{\textsc{Paper~III}}
\defcitealias{paper4}{\textsc{Paper~IV}}

\newcommand*{\vcenteredhbox}[1]{\begingroup
\setbox0=\hbox{#1}\parbox{\wd0}{\box0}\endgroup}

\begin{document}
{\hfill FERMILAB-PUB-25-0065-LDRD-PPD}
\title{The DECADE cosmic shear project III: validation of analysis pipeline using spatially inhomogeneous data}
\shortauthors{Anbajagane \& Chang et. al}
\shorttitle{The DECADE cosmic shear project III: analysis methodology and validation}

\author{\OrcidIDName{0000-0003-3312-909X}{D. Anbajagane}
(\vcenteredhbox{\includegraphics[height=1.2\fontcharht\font`\B]{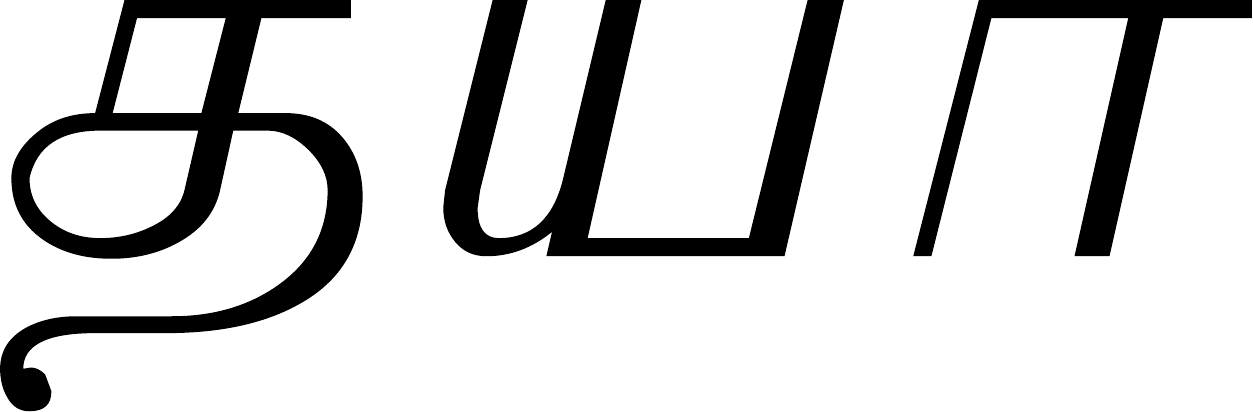}})$^\star$}
\affiliation{Department of Astronomy and Astrophysics, University of Chicago, Chicago, IL 60637, USA}
\affiliation{Kavli Institute for Cosmological Physics, University of Chicago, Chicago, IL 60637, USA}
\email{$^{\star}$dhayaa@uchicago.edu, chihway@kicp.uchicago.edu}

\author{\OrcidIDName{0000-0002-7887-0896}{C.~Chang}$^\star$}
\affiliation{Department of Astronomy and Astrophysics, University of Chicago, Chicago, IL 60637, USA}
\affiliation{Kavli Institute for Cosmological Physics, University of Chicago, Chicago, IL 60637, USA}

\author{\OrcidIDName{0009-0005-1143-495X}{N.~Chicoine}}
\affiliation{Department of Astronomy and Astrophysics, University of Chicago, Chicago, IL 60637, USA}
\affiliation{Department of Physics and Astronomy, University of Pittsburgh, 3941 O’Hara Street, Pittsburgh, PA 15260}

\author{\OrcidIDName{0000-0002-6002-4288}{L.~F.~Secco}}
\affiliation{Kavli Institute for Cosmological Physics, University of Chicago, Chicago, IL 60637, USA}

\author{\OrcidIDName{0000-0003-0478-0473}{C.~Y.~Tan}}
\affiliation{Department of Physics, University of Chicago, Chicago, IL 60637, USA}
\affiliation{Kavli Institute for Cosmological Physics, University of Chicago, Chicago, IL 60637, USA}

\author{\OrcidIDName{0000-0001-6957-1627}{P.~S.~Ferguson}}
\affiliation{DIRAC Institute, Department of Astronomy, University of Washington, 3910 15th Ave NE, Seattle, WA, 98195, USA}

\author{\OrcidIDName{0000-0001-8251-933X}{A.~Drlica-Wagner}}
\affiliation{Fermi National Accelerator Laboratory, P. O. Box 500, Batavia, IL 60510, USA}
\affiliation{Department of Astronomy and Astrophysics, University of Chicago, Chicago, IL 60637, USA}
\affiliation{Kavli Institute for Cosmological Physics, University of Chicago, Chicago, IL 60637, USA}

\author{\OrcidIDName{0000-0003-4394-7491}{K.~Herron}}
\affiliation{Department of Physics and Astronomy, Dartmouth College, Hanover, NH 03755, USA}

\author{\OrcidIDName{0000-0002-6904-359X}{M.~Adamow}}
\affiliation{Center for Astrophysical Surveys, National Center for Supercomputing Applications, 1205 West Clark St., Urbana, IL 61801, USA}
\affiliation{Department of Astronomy, University of Illinois at Urbana-Champaign, 1002 W. Green Street, Urbana, IL 61801, USA}

\author{\OrcidIDName{0000-0002-4588-6517}{R.~A.~Gruendl}}
\affiliation{Center for Astrophysical Surveys, National Center for Supercomputing Applications, 1205 West Clark St., Urbana, IL 61801, USA}
\affiliation{Department of Astronomy, University of Illinois at Urbana-Champaign, 1002 W. Green Street, Urbana, IL 61801, USA}

\author{\OrcidIDName{0000-0001-7774-2246}{M.~R.~Becker}}
\affiliation{Argonne National Laboratory, 9700 South Cass Avenue, Lemont, IL 60439, USA}

\author{\OrcidIDName{0000-0002-5279-0230}{R.~Teixeira}}
\affiliation{Department of Astronomy and Astrophysics, University of Chicago, Chicago, IL 60637, USA}
\affiliation{Department of Physics, Duke University Durham, NC 27708, USA}

\author{\OrcidIDName{0000-0002-7523-582X}{Z.~Zhang}}
\affiliation{Department of Astronomy and Astrophysics, University of Chicago, Chicago, IL 60637, USA}
\affiliation{Department of Physics, Stanford University, 382 Via Pueblo Mall, Stanford, CA 94305, USA}
\affiliation{SLAC National Accelerator Laboratory, Menlo Park, CA 94025, USA}

\author{\OrcidIDName{0000-0001-8505-1269}{A.~Alarcon}}
\affiliation{Institute of Space Sciences (ICE, CSIC),  Campus UAB, Carrer de Can Magrans, s/n,  08193 Barcelona, Spain}

\author{\OrcidIDName{0000-0003-2911-2025}{D.~Suson}}
\affiliation{Department of Chemistry and Physics, Purdue University Northwest 2200, 169th Ave, Hammond, IN 46323}

\author{\OrcidIDName{0000-0002-3173-2592}{A.~N.~Alsina}}
\affiliation{Instituto de Física Gleb Wataghin, Universidade Estadual de Campinas, 13083-859, Campinas, SP, Brazil}

\author{\OrcidIDName{0000-0002-6445-0559}{A.~Amon}}
\affiliation{Department of Astrophysical Sciences, Princeton University, Peyton Hall, Princeton, NJ 08544, USA}

\author{\OrcidIDName{0000-0003-0171-6900}{F.~Andrade-Oliveira}}
\affiliation{Physik-Institut, University of Zurich, Winterthurerstrasse 190, CH-8057 Zurich, Switzerland}

\author{\OrcidIDName{0000-0002-4687-4657}{J.~Blazek}}
\affiliation{Department of Physics, Northeastern University, Boston, MA 02115, USA}

\author{\OrcidIDName{0000-0001-5871-0951}{H.~Camacho}}
\affiliation{Physics Department, Brookhaven National Laboratory, Upton, NY 11973}

\author{\OrcidIDName{0000-0002-3690-105X}{J.~A.~Carballo-Bello}}
\affiliation{Instituto de Alta Investigaci\'on, Universidad de Tarapac\'a, Casilla 7D, Arica, Chile}

\author{\OrcidIDName{0000-0003-1697-7062}{W.~Cerny}}
\affiliation{Department of Astronomy, Yale University, New Haven, CT 06520, USA}

\author{\OrcidIDName{0000-0003-1680-1884}{Y.~Choi}}
\affiliation{NSF National Optical-Infrared Astronomy Research Laboratory, 950 North Cherry Avenue, Tucson, AZ 85719, USA}

\author{\OrcidIDName{0000-0003-4480-0096}{C.~Doux}}
\affiliation{Université Grenoble Alpes, CNRS, LPSC-IN2P3, 38000 Grenoble, France}

\author{\OrcidIDName{0000-0001-6134-8797}{M.~Gatti}}
\affiliation{Kavli Institute for Cosmological Physics, University of Chicago, Chicago, IL 60637, USA}

\author{\OrcidIDName{0000-0003-3270-7644}{D.~Gruen}}
\affiliation{University Observatory, Faculty of Physics, Ludwig-Maximilians-Universität, Scheinerstr. 1, 81679 Munich, Germany}
\affiliation{Excellence Cluster ORIGINS, Boltzmannstr. 2, 85748 Garching, Germany}

\author{\OrcidIDName{0000-0001-5160-4486}{D.~J.~James}}
\affiliation{Applied Materials Inc., 35 Dory Road, Gloucester, MA 01930}
\affiliation{ASTRAVEO LLC, PO Box 1668, Gloucester, MA 01931}

\author{\OrcidIDName{0000-0001-8356-2014}{E.~Krause}}
\affiliation{Department of Astronomy/Steward Observatory, University of Arizona, Tucson, AZ 85721 USA}

\author{\OrcidIDName{0000-0003-2511-0946}{N.~Kuropatkin}}
\affiliation{Fermi National Accelerator Laboratory, P. O. Box 500, Batavia, IL 60510, USA}

\author{\OrcidIDName{0000-0002-9144-7726}{C.~E.~Mart\'inez-V\'azquez}}
\affiliation{International Gemini Observatory/NSF NOIRLab, 670 N. A'ohoku Place, Hilo, Hawai'i, 96720, USA}

\author{\OrcidIDName{0000-0002-8093-7471}{P.~Massana}}
\affiliation{NSF's NOIRLab, Casilla 603, La Serena, Chile}

\author{\OrcidIDName{0000-0003-3519-4004}{S.~Mau}}
\affiliation{Kavli Institute for Particle Astrophysics \& Cosmology, P.O.\ Box 2450, Stanford University, Stanford, CA 94305, USA}
\affiliation{Department of Physics, Stanford University, 382 Via Pueblo Mall, Stanford, CA 94305, USA}

\author{\OrcidIDName{0000-0002-4475-3456}{J.~McCullough}}
\affiliation{Department of Astrophysical Sciences, Peyton Hall, Princeton University, Princeton, NJ USA 08544}

\author{\OrcidIDName{0000-0003-0105-9576}{G.~E.~Medina}}
\affiliation{David A. Dunlap Department of Astronomy \& Astrophysics, University of Toronto, 50 St George Street, Toronto ON M5S 3H4, Canada}
\affiliation{Dunlap Institute for Astronomy \& Astrophysics, University of Toronto, 50 St George Street, Toronto, ON M5S 3H4, Canada}

\author{\OrcidIDName{0000-0001-9649-4815}{B.~Mutlu-Pakdil}}
\affiliation{Department of Physics and Astronomy, Dartmouth College, Hanover, NH 03755, USA}

\author{\OrcidIDName{0000-0001-9438-5228}{M. ~ Navabi}}
\affiliation{Department of Physics, University of Surrey, Guildford GU2 7XH, UK}

\author{\OrcidIDName{0000-0002-8282-469X}{N.~E.~D.~Noël}}
\affiliation{Department of Physics, University of Surrey, Guildford GU2 7XH, UK}

\author{\OrcidIDName{0000-0002-6021-8760}{A.~B.~Pace}}
\affiliation{Department of Astronomy, University of Virginia, 530 McCormick Road, Charlottesville, VA 22904, USA}

\author{\OrcidIDName{0000-0002-2762-2024}{A.~Porredon}}
\affiliation{Centro de Investigaciones Energ\'eticas, Medioambientales y Tecnol\'ogicas (CIEMAT), Madrid, Spain}

\author{\OrcidIDName{0000-0002-7354-3802}{M.~Raveri}}
\affiliation{Department of Physics and INFN, University of Genova, Genova, Italy}

\author{\OrcidIDName{0000-0001-5805-5766}{A.~H.~Riley}}
\affiliation{Institute for Computational Cosmology, Department of Physics, Durham University, South Road, Durham DH1 3LE, UK}

\author{\OrcidIDName{0000-0002-1594-1466}{J.~D.~Sakowska}}
\affiliation{Department of Physics, University of Surrey, Guildford GU2 7XH, UK}

\author{\OrcidIDName{0000-0001-7147-8843}{S.~Samuroff}}
\affiliation{Institut de F\'{i}sica d'Altes Energies, The Barcelona Institute of Science and Technology, Campus UAB, 08193 Bellaterra (Barcelona) Spain}

\author{\OrcidIDName{0000-0003-3054-7907}{D.~Sanchez-Cid}}
\affiliation{Physik-Institut, University of Zurich, Winterthurerstrasse 190, CH-8057 Zurich, Switzerland}
\affiliation{Centro de Investigaciones Energéticas, Medioambientales y Tecnológicas (CIEMAT), Madrid, Spain}

\author{\OrcidIDName{0000-0003-4102-380X}{D.~J.~Sand}}
\affiliation{Steward Observatory, University of Arizona, 933 North Cherry Avenue, Tucson, AZ 85721-0065, USA}

\author{\OrcidIDName{0000-0003-3402-6164}{L.~Santana-Silva}}
\affiliation{Centro Brasileiro de Pesquisas F\'isicas, Rua Dr. Xavier Sigaud 150, 22290-180 Rio de Janeiro, RJ, Brazil}

\author{\OrcidIDName{0000-0001-6082-8529}{M.~Soares-Santos}}
\affiliation{Physik-Institut, University of Zurich, Winterthurerstrasse 190, CH-8057 Zurich, Switzerland}

\author{\OrcidIDName{0000-0003-1479-3059}{G.~S.~Stringfellow}}
\affiliation{Center for Astrophysics and Space Astronomy, University of Colorado, 389 UCB, Boulder, CO 80309-0389, USA}

\author{\OrcidIDName{0000-0001-7836-2261}{C.~To}}
\affiliation{Kavli Institute for Cosmological Physics, University of Chicago, Chicago, IL 60637, USA}

\author{\OrcidIDName{0000-0003-4341-6172}{A.~K.~Vivas}}
\affiliation{Cerro Tololo Inter-American Observatory/NSF NOIRLab, Casilla 603, La Serena, Chile}

\author{\OrcidIDName{0000-0003-1585-997X}{M.~Yamamoto}}
\affiliation{Department of Astrophysical Sciences, Princeton University, Peyton Hall, Princeton, NJ 08544, USA}

\author{\OrcidIDName{0000-0001-6455-9135}{A.~Zenteno}}
\affiliation{Cerro Tololo Inter-American Observatory/NSF NOIRLab, Casilla 603, La Serena, Chile}

\author{\OrcidIDName{0000-0001-9789-9646}{J.~Zuntz}}
\affiliation{Institute for Astronomy, University of Edinburgh, Edinburgh EH9 3HJ, UK}

\begin{abstract}
We present the pipeline for the cosmic shear analysis of the Dark Energy Camera All Data Everywhere (DECADE) weak lensing dataset: a catalog consisting of 107 million galaxies observed by the Dark Energy Camera (DECam) in the northern Galactic cap. The catalog derives from a large number of disparate observing programs and is therefore more inhomogeneous across the sky compared to existing lensing surveys. First, we use simulated data-vectors to show the sensitivity of our constraints to different analysis choices in our inference pipeline, including sensitivity to residual systematics. Next we use simulations to validate our covariance modeling for inhomogeneous datasets. Finally, we show that our choices in the end-to-end cosmic shear pipeline are robust against inhomogeneities in the survey, by extracting relative shifts in the cosmology constraints across different subsets of the footprint/catalog and showing they are all consistent within $1\sigma$ to $2\sigma$. This is done for forty-six subsets of the data and is carried out in a fully consistent manner: for each subset of the data, we rederive the photometric redshift estimates, shear calibrations, survey transfer functions, the data vector, measurement covariance, and finally, the cosmological constraints. Our results show that existing analysis methods for weak lensing cosmology can be fairly resilient towards inhomogeneous datasets. This also motivates exploring a wider range of image data for pursuing such cosmological constraints. 
\end{abstract}


\section{Introduction}

Over the past two decades, weak gravitational lensing (also referred to as weak lensing or cosmic shear) has emerged as a leading probe in constraining the cosmological parameters of our Universe \citep{Asgari2021, Secco2021, Amon2021, Dalal2023, Wright:2025:KidsCosmo}. Weak lensing refers to the subtle bending of light from distant ``source galaxies'' due to the large-scale matter distribution between the source and the observer \citep{Bartelmann2001}. Thus, weak lensing, through its sensitivity to the matter distribution, probes the large-scale structure (LSS) of our Universe and any phenomenon that impact this structure; including cosmological ones such as modified gravity \citep[\eg][]{Schmidt:2008:MG_WL} and primordial signatures \citep[\eg][]{Anbajagane2023Inflation, Goldstein:2024:inflation, Primordial1, Primordial2}, as well as a wide variety of astrophysical ones \citep[\eg][]{Chisari2018BaryonsPk, Schneider2019Baryonification, Arico:2021:Bacco, Grandis:2024:XrayLensing, Bigwood:2024:BaryonsWLkSZ}. Weak lensing has many advantages in the landscape of cosmological probes, the primary of which is that it is an \textit{unbiased} tracer of the density field  --- unlike other tracers, such as galaxies --- and does not require modeling or marginalizing over an associated bias parameter \citep{Bartelmann2001}. Since the lensing signal is also sensitive to the distance between the source galaxy, the intervening matter, and the observer, lensing is also a probe of the geometry of our Universe. For these reasons, it is one of the leading probes of cosmology and has delivered some of our best constraints on cosmological parameters.

This paper is part of a series of works detailing the \decade cosmic shear analysis. \href{\#cite.paper1}{Anbajagane \& Chang et al. \citeyear{paper1}} (hereafter \citetalias{paper1}) describes the shape measurement method, the derivation of the final cosmology sample, the robustness tests, and also the image simulation pipeline from which we quantify the shear calibration uncertainty of this sample. \citet[][hereafter \citetalias{paper2}]{paper2} derives both the tomographic bins and calibrated redshift distributions for our cosmology sample, together with a series of validation tests. This work (\textsc{Paper III}) describes the methodology and validation of the model, in addition to a series of survey inhomogeneity tests. Finally \href{\#cite.paper4}{Anbajagane \& Chang et al. \citeyear{paper4}} (hereafter \citetalias{paper4}) shows our cosmic shear measurements and presents the corresponding constraints on cosmological models.

This work serves three, key purposes. First, to detail the modeling/methodology choices of the cosmic shear analysis, and the robustness of our results to said choices. Second, to build on the null-tests of \citetalias{paper1} and show that our data vector (and cosmology) are not susceptible to contamination from systematic effects, such as correlated errors in the point-spread function (PSF) modeling. Finally, we show that the end-to-end cosmology inference pipeline is resilient to the impact of spatial inhomogeneities. As highlighted in both \citetalias{paper1} and \citetalias{paper2}, the \decade dataset contains some unique characteristics relative to other weak lensing datasets; particularly, the spatial inhomogeneity in the image data coming from this dataset's origin as an amalgamation of many different public observing programs. We perform a suite of tests where we rerun the end-to-end pipeline for different subsets of our data --- where each subset contains specific kinds of galaxies (red/blue, faint/bright, \textit{etc}.) or contains objects measured in regions of the sky with better/worse image quality (changes in seeing, airmass, interstellar extinction, \textit{etc}.) --- and show that our cosmology constraints are robust across such subsets. 

This paper is structured as follows. In Section \ref{sec:data}, we briefly describe the \decade shape catalog, and in Section~\ref{sec:model}, we present the cosmology model used in the \decade cosmic shear project. In Section~\ref{sec:inference}, we outline the different components required for parameter inference, including our analytic covariance matrix. In Section~\ref{sec:validation}, we check the robustness of our constraints across modeling choice in simulated data vectors. Section~\ref{sec:SpatialInhomog} details our tests on the sensitivity of our parameter constraints to spatial inhomogeneity and to different selections of the source galaxy catalog. We conclude in Section \ref{sec:summary}.

\section{The DECADE dataset}\label{sec:data}

The \decade dataset is a galaxy shape catalog of 107 million galaxies, spanning $5,\!412 \deg^2$ of the sky. The catalog is introduced in \citetalias{paper1}, alongside a suite of null-tests and shear calibrations made using image simulations of the survey data. The galaxy shears are estimated using the \textsc{Metacalibration} method \citep{Sheldon:2017:Shear}, with an approach designed to mimic that of DES Y3 \citep{y3-shapecatalog}. In \citetalias{paper2}, we split the catalog into four tomographic bins and estimate the redshift distribution of the ensembles in each bin using the self-organizing maps photometric redshifts (SOMPZ) method \citep{Buchs2019, Myles:2021:DESY3}, including cross-checking these estimates with alternative methods such as clustering redshifts \citep{Schneider:2006:WZ,Newman:2008:WZ,Menard2013, Davis2017, Gatti:2018:WZ, Gatti:2022:WzY3}.

\begin{figure*}
    \centering
    \includegraphics[width=2\columnwidth]{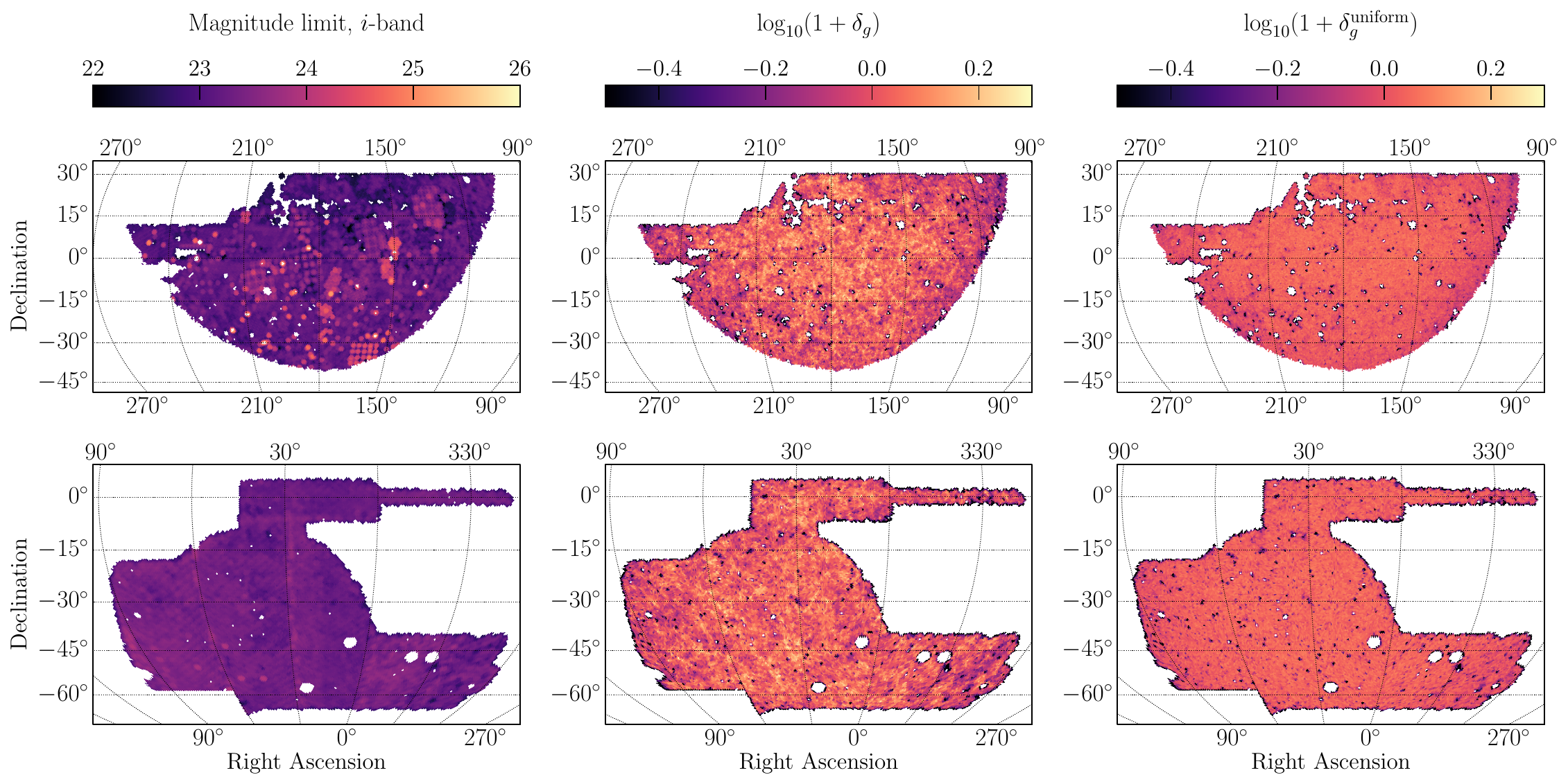}
    \caption{A set of maps from the \decade (top) and DES Y3 (bottom) datasets. From left to right, we show the (apparent) magnitude limit map in the $i$-band (estimated at $S/N = 10$ within a $2\arcsec$ aperture; see \citetalias{paper1}), the overdensity in source galaxy number counts $\delta_g$, and the same after randomizing the galaxy positions within the survey mask. All maps are shown at $\texttt{NSIDE}=1024$. While the \decade magnitude limit map has significantly more structure relative to that of DES Y3, the \decade source-galaxy sample has much less inhomogeneity as the \decade and DES Y3 source-galaxy samples are defined with certain cuts (specifically $m_i < 23.5$, where $m_i$ is the $i$-band magnitude of the source galaxy) that make the variation in $\delta_g$ similar across the two surveys. The remaining large-scale variations, found towards the eastern/western edges of the footprint, are due to the closer proximity of the \decade footprint to the Galactic plane. The right-most panels show, as a simple reference, the number-density fluctuations for a sample with randomized positions. There are still some small fluctuations due to the mask structure on small scales; the mask is defined at $\texttt{NSIDE}=4096$ \citepalias{paper1}.}
    \label{fig:Survey_homogeneity}
\end{figure*}

The \decade dataset is constructed from available community data (collected up to December 2022) from the Dark Energy Camera \citep[DECam,][]{Flaugher2015b} within the footprint of interest to us; see Figure 2 in \citetalias{paper1} for the survey footprint, and also Section 2.1 in that work. The nature of this dataset --- as an amalgamation of available archival data, rather than as a dedicated cosmological survey program --- results in significant inhomogeneities in the survey observing conditions (such as depth). This, in turn, propagates into the observed object properties and therefore, the detection/selection functions of galaxy samples. While such inhomogeneities in observing conditions exist for other photometric galaxy surveys, such as DES, the variations are smaller relative to the \decade inhomogeneities; see Figure~\ref{fig:Survey_homogeneity} (which is discussed in more detail in Section \ref{sec:cov}). The latter part of this work (Section \ref{sec:SpatialInhomog}) explicitly shows the consistency in cosmology constraints when limiting our dataset to different regions of the sky with more/less spatial inhomogeneity in the chosen image quality.

In general, the constraining power of our catalog is fairly similar to that of DES Y3 \citepalias{paper1}. We have a source-galaxy number density that is $\approx\!20\%$ lower than that of DES Y3 ($n_{\rm eff} = 4.6\,\, \text{arcmin}^{-2}$ for \decade compared to $n_{\rm eff} = 5.6 \,\, \text{arcmin}^{-2}$ in DES Y3\footnote{Here, $n_{\rm eff}$ is the effective source-galaxy number density that takes into account the galaxy weight. See \citetalias{paper1} for more details on how these weights are generated.}); however, our footprint is $\approx\! 25\%$ larger ($5412 \deg^2$ vs. $4143 \deg^2$). Thus, our lower number density is compensated by our larger area, and our shape catalog's statistical precision is similar to that of DES Y3 \citep{y3-shapecatalog}. Note, however, that our sample has a slightly lower mean redshift ($\Delta \langle z \rangle \approx 0.02$), particularly in the highest tomographic bins \citepalias{paper2}.

\section{modeling}\label{sec:model}

Our modeling pipeline follows the approach of \citet{Krause2021}, who detailed the pipeline for the DES Y3 analyses. We reproduce the salient parts of the model here, and direct the interested reader to that work for a more detailed discussion. We use the two-point correlation function of galaxy shapes, $\xi_{\pm}(\theta)$ \citep{Bartelmann2001}, computed for different angular separations $\theta$, as our primary cosmological measurement. We have
\begin{align} \label{eq:xipm}
    \xi_{\pm}^{ij}(\theta) = \sum_{\ell} & \,\,\frac{2\ell+1}{2\pi\ell^{2}\left(\ell+1\right)^{2}}\left[G_{\ell,2}^{+}\left(\cos\theta\right)\pm G_{\ell,2}^{-}\left(\cos\theta\right)\right]\nonumber\\& \times\left[C_{EE}^{ij}(\ell)\pm C_{BB}^{ij}(\ell)\right],
\end{align}
\noindent
where $\ell$ is the harmonic multipole, the functions $G^\pm_\ell(x)$ are computed from Legendre polynomials $P_\ell(x)$ and averaged over angular bins \citep{Krause2021}, and the $i$ and $j$ indices specify the two tomographic redshift bins from which the correlation function is calculated. Assuming the extended Limber approximation \citep{Limber1953,Loverde2008} and in a spatially flat universe, the lensing power spectrum can be written as 
\begin{equation}
    C^{ij}_{EE}(\ell) = \int_{0}^{\chi_{\rm H}} d\chi \frac{q^{i}(\chi)q^{j}(\chi)}{\chi^2} P_{\rm NL}\left( \frac{\ell + 1/2}{\chi}, \chi \right),
\label{eq:Cl}
\end{equation}
where $\chi$ is the radial comoving distance, $\chi_{\rm H}$ is the distance to the horizon, $P_\text{NL}$ is the nonlinear matter power spectrum, and $q(\chi)$ is the lensing efficiency defined via
\begin{equation}
    q^{i}(\chi) = \frac{3}{2} \Omega_{\rm m} \left( \frac{H_{0}}{c}\right)^{2} \frac{\chi}{a(\chi)} \int_{\chi}^{\chi_{H}}d\chi ' n_{i}(\chi') \frac{\chi' - \chi}{\chi'},
\label{eq:lensing_efficiency}
\end{equation}
where $\Omega_{\rm m}$ is the matter density today, $H_{0}$ is the Hubble parameter today, $a$ is the scale factor, $c$ is the speed of light, and $n_{i}(\chi)$ is the redshift distribution of the bin $i$. The $B$-mode angular power spectrum, $C^{ij}_{BB}$, will be non-zero due to the presence of intrinsic alignments (IA) of galaxy shapes \citep[\eg][]{TroxelIshak2014, Blazek2019};\footnote{In practice, lensing also generates a $C^{ij}_{BB}$ term but its amplitude is orders of magnitude lower than the $C^{ij}_{EE}$ term and is therefore ignored \citep[\eg][]{Krause:2010:Bmodes}.} see Section \ref{sec:model:IA} for discussions on our IA model.

Following the model described above, we consider several additional elements that are briefly described below. We note that we have largely followed the modeling framework used in the DES Year 3 (Y3) analysis \citep{Krause2021}, which is implemented in the public version of the \textsc{CosmoSIS} software \citep{Zuntz2015}. We will explicitly identify cases where our model deviates from that of DES Y3. 

\subsection{Nonlinear power spectrum}\label{sec:model:Pk_Nl}

On small scales, the nonlinearity of structure formation as induced by gravity necessitates a modification of the matter power spectrum from the linear model, where the latter is typically calculated via a Boltzmann solver such as \textsc{CAMB} \citep{Lewis2000} or \textsc{Class} \citep{Lesgourgues:2011:CLASS}. As the resolution and fidelity of $N$-body simulations have improved, so too have prescriptions for the nonlinear power spectrum. In the DES Y1 and Y3 cosmic shear analysis, \textsc{HaloFit} \citep{Takahashi2012} was used and validated as the default nonlinear power spectrum. However, the study and joint-analysis between DES and the Kilo Degree Survey (KiDS) in \citet{DESKiDS2023} show that the updated version of \textsc{HMCode} \citep[\textsc{HMCode2020},][]{Mead2020a, Mead2021b} is more consistent with the $N$-body simulation-based Euclid emulator \citep{Euclid2019}. At $k<10h {\rm Mpc}^{-1}$, \textsc{HMCode2020} has an accuracy of $<2.5\%$ when compared to $N$-body simulations. As a result, we adopt \textsc{HMCode2020} as the fiducial model for our nonlinear power spectrum. Note that for the constraining power of the \decade data, we do not expect any significant differences in constraints when switching between the \textsc{HaloFit} or \textsc{HMCode2020} models. We explicitly test this modeling choice in our simulated chains below, and also check it directly on the data in \citetalias{paper4}.

The above discussion pertains to just the impact of gravitationally sourced nonlinear evolution in the matter power spectrum. The structure on small scales is also impacted by the non-gravitational processes of baryons; most prominently, the ejection of gas from the halo due to feedback processes \citep{Chisari2018BaryonsPk}. Hydrodynamical simulations \citep[see][for a review]{Vogelsberger2020Hydro}, which model these effects through approximate subgrid prescriptions, generate a variety of predictions for the properties of halos in simulations \citep[\eg][]{Anbajagane2020StellarProp, Lim2021GasProp, Lee2022rSZ, Stiskalek2022TNGHorizon, Anbajagane2022Baryons, Anbajagane2022GalaxyVelBias, Shao2022Baryons, Shao2023Baryons, Gebhardt2023CamelsAGNSN} and therefore also for the changes in the nonlinear matter power spectrum \citep[\eg][]{vanDaalen2011, Springel:2018:FirstTNG, Chisari2018BaryonsPk, Amon2022S8Baryons, Salcido:2023:SPk}. These variations have been captured through phenomenological, halo-based models such as ``baryonification'' \citep{Schneider:2015:Baryons, Schneider2019Baryonification, Arico:2021:Bacco, Anbajagane:2024:Baryonification} and also a variant of the \textsc{HMCode2020} model discussed above, denoted as ``\textsc{HMCode2020} Feedback'' (following the nomenclature used in \textsc{CAMB}).

The ``\textsc{HMCode2020} Feedback'' model includes a $T_{\rm AGN}$ parameter (here, AGN stands for Active Galactic Nuclei) --- which can be interpreted as an ``effective feedback'' parameter with larger values indicating stronger suppression of matter clustering --- that quantifies the baryonic contribution to the nonlinear matter power spectrum, and this parameter was varied in the joint-analysis of DES and KiDS \citep{DESKiDS2023}. In this work, our fiducial analysis does not use any of these techniques and follows DES Y3 \citep{Secco2021, Amon2021}: we remove any scales where the baryonic effects are significant and analyze the remaining scales with a model that does not account for baryons. The identification of scales to be removed is done using the procedure described in Section~\ref{sec:scalecuts}. However, we also perform a variant analysis in \citetalias{paper4} that uses all measurements (\ie does not perform any scale cuts) and instead varies the $T_{\rm AGN}$ parameter to include the effects of baryons.

\subsection{Intrinsic alignment (IA)}\label{sec:model:IA}

While cosmic shear manifests in the spatial correlations of galaxy shapes, these shapes can also be spatially correlated in the absence of lensing due to the gravitational environment, or tidal fields, that the galaxies form within \citep[see, \eg the reviews of ][and references therein]{TroxelIshak2014, Kiessling:2015:IA, Lamman:2024:IA}. This effect is referred to as intrinsic alignment. The IA signal is directly connected to the matter distribution and therefore can be predicted using the matter power spectrum.

The DES Y3 cosmic shear analysis uses the Tidal Alignment and Tidal Torquing model \citep[TATT,][]{Blazek2019}, which  allows for five free nuisance parameters: a tidal alignment amplitude ($a_1$) with redshift evolution ($\eta_1$); a tidal torquing amplitude ($a_2$) with redshift evolution ($\eta_2$); and a linear bias amplitude ($b_{\rm TA}$). The TATT model reduces to the simpler, Nonlinear linear alignment model \citep[NLA,][]{Hirata2004,Bridle2007,Joachimi2011} when setting $a_1=A_{\rm IA}$, $\eta_1=\eta$, and $a_2=\eta_2=b_{\rm TA}=0$. Different analyses of DES Y3 have used either TATT \citep{Secco2021,Amon2021, Samuroff2022} or NLA \citep{DESKiDS2023} as their fiducial choice. For the \decade cosmic shear analysis, we adopt the more conservative TATT model as our fiducial choice but test the change in our results if we revert to the simpler NLA model.\footnote{Since TATT has more parameters, it may allow the fit to explore unphysical or weakly constrained regions of parameter space. Thus, results are often cross-checked using the simpler NLA model.}

The amplitude of the IA contribution is parameterized as,
\begin{align}\label{eqn:IA_ampl}
    A_1(z) &= -a_1 \bar{C}_1 \frac{\rho_{\text{crit}} \Om}{D(z)} 
    \left( \frac{1 + z}{1 + z_0} \right)^{\eta_1}, \\
    A_2(z) &= 5a_2 \bar{C}_1 \frac{\rho_{\text{crit}} \Om}{D^2(z)} 
    \left( \frac{1 + z}{1 + z_0} \right)^{\eta_2},\\
    A_{1\delta}(z) &= b_{\rm TA} A_1(z),
\end{align}
where $A_1$ and $A_2$ scale the matter power spectra, $D(z)$ is the linear growth rate, $\rho_{\text{crit}}$ is the critical density at $z = 0$, and $\bar{C}_1 = 5 \times 10^{-14} M_\odot h^{-2} \text{Mpc}^2$ is a normalization constant, set by convention. We choose a pivot redshift, $z_0 = 0.62$ following \citet{Secco2021, Amon2021}. The free parameters of our model are the amplitudes $a_1, a_2, b_{\rm TA}$ and the power-law indices $\eta_1, \eta_2$. As mentioned above, the NLA model is obtained by setting $a_2, \eta_2, b_{\rm TA} = 0$. See Equations 20--23 in \citet{Secco2021} for a description of the different IA-related power-spectra that contribute to the final signal.

\subsection{Shear and redshift calibration uncertainties}\label{sec:model:Calib}

The two main sources of systematic uncertainties on our measurements are those associated with estimates of the shear and of the redshift distribution. We follow the implementation from DES Y3 \citep{Secco2021, Amon2021} --- which is consistent with those of numerous other works \citep[\eg][]{Asgari2021, Dalal2023} --- in the modeling of these uncertainties. We now detail our implementation below.

For any shear measurement algorithm, a calibration step is needed for the measured shear. This calibration, which is obtained either from simulations or the data itself, carries an associated uncertainty. In the weak lensing regime, the residual impact of any biases on the measured shear, $\gamma_{\textrm{obs}}$, is modeled with both multiplicative $m_{i}$ and additive components $c_{i}$ \citep{Huterer2006,Heymans2006,Bridle2007}, that scale the true shear $\gamma_{\textrm{true}}$ as
\begin{equation}
    \gamma_{\textrm{obs}} = \gamma_{\textrm{true}} (1+m_{i}) + c_{i},
\end{equation}
where $m_i$ and $c_i$ are constants corresponding to the $i^{\rm th}$ redshift bin. The additive, residual bias often has a sufficiently small amplitude to not be relevant \citep[\eg][]{Macrann2022ImSim, Amon2021}. Furthemore, the shear measurements of our catalog are mean-subtracted \citepalias{paper1} following the approach in other works \citep[\eg][]{Zuntz2018, y3-shapecatalog, Yamamoto2025}. This naturally corrects for such additive biases in the data. The multiplicative residual bias, however, scales the amplitude of $\xi_{\rm \pm}$ and thus propagates directly to key cosmological constraints. As a result, accurate estimates of $m_i$ are a necessity. We have made such estimates for our shear catalog using a suite of simulated images of the \decade survey, as described in detail in \citetalias{paper1}. The simulations inform priors for $m_i$ which are listed in Table~\ref{tab:params}. We note that these simulations are data-driven and reproduce the range of observing conditions seen in the real data; see \citetalias{paper1}.

For the redshift distribution, $n(z)$, we also follow DES Y3 in using a shift in the mean redshift, $\Delta z_{i}$, per bin that then shifts the measured $n(z)$ for each bin $i$ as 
\begin{equation}
    n_{i}(z) = n_{{\rm orig},i}\left(z - \Delta z_{i}\right).
\end{equation}
where $n_{{\rm orig},i}$ is the original distribution. This parametrization keeps the shape of the $n(z)$ unchanged. While one could explore more complicated parametrizations of the $n(z)$ uncertainties, the above model is sufficient for the statistical power of our data (which is similar to DES Y3). In \citetalias{paper2}, we derived the priors on $\Delta z_{i}$ using the self-organizing map photometric redshifts (SOMPZ) method while accounting for a variety of uncertainty contributions including cosmic variance, shot noise, redshift bias, and photometric zeropoint uncertainties. In the Appendix A of \citetalias{paper2}, we also derived estimates of the $n(z)$ from combining SOMPZ with an independent estimate, clustering redshifts (WZ). The latter method includes information from the spatial clustering between our shear sample and spectroscopic sources. The former, SOMPZ-only method is our fiducial choice, while we use the latter SOMPZ plus WZ method as a cross-check. The calibrated priors on $\Delta z_i$ from the fiducial setup are listed in Table~\ref{tab:params}. Our redshift distributions generally span $0 < z < 1.5$. See Figure 9 of \citetalias{paper2} for more details.

\section{Parameter inference}
\label{sec:inference}

We now describe the components, in addition to the model described in Section~\ref{sec:model}, that are required for the final parameter inference.

\subsection{Scale cuts}
\label{sec:scalecuts}

As we discussed in Section~\ref{sec:model:Pk_Nl}, baryons can influence the matter power spectrum on a wide range of scales, presenting a significant source of uncertainty. For example, feedback processes from AGN and supernovae can heat environments and suppress matter clustering, while metal enrichment may offer cooling channels that increase small-scale clustering \citep[\eg][]{Chisari2018BaryonsPk, Schneider2019Baryonification}. To reduce the impact of these baryonic imprints, we follow \citet{Secco2021} and \citet{Amon2021} in removing shear measurements whose signal can have a significant contribution from baryon evolution. In particular, we use a simulated data vector which is generated using the matter-power spectrum measured in the ``AGN'' model of the OverWhelmingly Large Simulations \citep[OWLS,][see their Table 2]{Schaye:2010:OWLS}. The current estimates of the power-spectrum suppression from data are statistically consistent with this OWLS model \citep[\eg][see their Figure 1 and 4]{Bigwood:2024:BaryonsWLkSZ}.

The $\xi_\pm$ measurements used in this project are made in 20 logarithmically spaced bins over the angular scale range of $2.5\arcmin < \theta < 250 \arcmin$, following the choices of \citet{Secco2021} and \citet{Amon2021}. The scale cuts for these measurements are set by requiring the contaminated and uncontamined data vectors to differ by $\Delta \chi^2 < 0.3$. This roughly follows the scale-cuts procedure used in DES Y3 \citep{Secco2021}. 
In this work, we derive the scale cuts with the following steps, starting from the unmasked data vector, $D$:
\begin{itemize}
    \item Compute $(\Delta \chi^2)_i = (D^{i}_{\rm contam} - D^{i}_{\rm uncontam})^2/\sigma_i^2$ on each data point $i$ individually, where $\sigma_i$ is the uncertainty associated with data point $i$,\vspace{5pt}
    \item Find the measurement at the smallest scale in each bin. We have 10 combinations of tomographic bin pairs and two correlation functions, $\xi_+$ and $\xi_-$, resulting in 20 points selected for potential masking,\vspace{5pt}
    \item Of the 20 selected points from the previous step, we find the data point with the highest $\Delta \chi^2$ and remove it. We repeat this iteratively until the $\Delta \chi^2$ between the contaminated and uncontaminated data vectors, after masking/scale cuts, is $\leq 0.3$. This $\Delta \chi^2$ between the two data vectors is determined using the entire covariance matrix (and not just its diagonal component).
\end{itemize}
The $\xi_-$ shear measurements are more heavily affected by baryon contamination and thus require more impactful scale cuts. In practice, we remove a similar number of points as the fiducial DES Y3 scale cuts \citep{Secco2021, Amon2021}; our final data vector has 220 data points whereas the fiducial data vector from DES Y3 has 222 data points. The unmasked datavector has 400 points --- 20 angular separations per tomographic bin and per $\xi_+$ and $\xi_-$. Our data vector, and the associated scale cuts per tomographic bin, is shown in Figure 3 of \citetalias{paper4}.

As a final step, we analyze a data vector \textit{with} baryon imprints while using our fiducial model, which includes scale cuts but no baryon modelling. We then check that the cosmological constraints in the $\Omega_{\rm m} - S_8$ plane are shifted by less than $0.3 \sigma$. This result is detailed further below in Section~\ref{sec:scale_s8}. Note that there are other valid approaches for determining which measurements to discard from the full data vector. The only clear requirement is that after scale cuts are applied, the posterior constraint from the data vector must shift by less than $0.3\sigma$ when the data vector is contaminated with baryonic imprints. Our method above is one such approach and results in scale cuts that are similar to those of DES Y3, as mentioned above.

\subsection{Covariance matrix}
\label{sec:cov}

The parameter inference described in Section \ref{sec:likelihood} below requires estimating the covariance of our measurements. We now describe our approach to generating this covariance, and also check that the fiducial framework is sufficient for inhomogeneous datasets of the kind considered in this work.

In this work, we generate an analytical covariance using the \mbox{\cosmocov} package \citep{Krause:2017:CosmoLike, Fang2020, Fang2020b}. See references therein for the analytic formalism and validations of this formalism. There are many benefits to using an analytical covariance matrix over one estimated from simulations or from data resampling (e.g., jackknife covariances). A key advantage is that such a covariance matrix does not contain numerical noise and can therefore be robustly inverted to obtain the likelihood. This approach of analytically modeling the covariance is also adopted in the DES Y3 analyses \citep{Friedrich:2021:CovY3}. 

\cosmocov uses a two-dimensional \textsc{FFTLog} method \citep{Hamilton:2000:FFTLog} to efficiently calculate the real-space, non-Gaussian covariance matrix for galaxy weak lensing (as well as other probes not considered in this work). This covariance can be decomposed as,
\begin{equation}\label{eqn:cov:contributions}
    \mathbf{C} = \mathbf{C}_{\rm G} + \mathbf{C}_{\rm SSC} + \mathbf{C}_{\rm cNG},
\end{equation}
where $\mathbf{C}_{\rm G}$ is the standard Gaussian covariance matrix, and the two non-Gaussian contributions include a connected four-point, non-Gaussian correlation ($\mathbf{C}_{\rm cNG}$) that accounts for the presence of nonlinear structure \citep{Wagner:2015:Response, Barreira:2017:Response, Barreira:2017:ResponseCov} and then a super sample contribution ($\mathbf{C}_{\rm SSC}$) which accounts for cross-correlations between different small-scale modes induced by modes larger than the survey footprint \citep[\eg][]{Barreira:2018:SSC}. The non-Gaussian contributions are subdominant for the diagonal elements of $\mathbf{C}$ but are dominant in the off-diagonal terms as their primary effect is in coupling together different modes/scales. 

Certain terms in the final covariance can have some sensitivity to the geometry of the survey mask, and some can also have sensitivity to the distribution of galaxies within that mask (we detail the exact terms further below). We first recast Equation \eqref{eqn:cov:contributions} --- to distinguish the contributions from noise and those from cosmic variance (CV) --- by splitting the covariance into the following terms
\begin{equation}
    \mathbf{C} = \mathbf{C}_{\rm SN} + \mathbf{C}_{\rm mix} + \mathbf{C}_{\rm CV/SSC/cNG},
\end{equation}
where $\mathbf{C}_{\rm SN}$ is the shape-noise term, $\mathbf{C}_{\rm CV/SSC/cNG}$ is the sum of the cosmological terms (\ie the sole term in the covariance when shape noise is negligible), and $\mathbf{C}_{\rm mix}$ is the mixed term which contains one power each of the cosmological signal and the noise field. In \cosmocov, the terms $\mathbf{C}_{\rm SN}$ and $\mathbf{C}_{\rm SSC}$ (that is, just the super-sample covariance term in $\mathbf{C}_{\rm CV/SSC/cNG}$) include the effects of the survey mask, while the $\mathbf{C}_{\rm CV}$ and $\mathbf{C}_{\rm cNG}$ terms ignore it. \citet{Friedrich:2021:CovY3} explicitly show that the impact of the mask on the mixed covariance term, as well as the cosmic variance and connected four-point terms, changes the posterior width by less than $1\%$ (see their Figure 2). The \decade survey footprint has a similarly regular geometry relative to that of the DES footprint,\footnote{The structure of our small-scale mask features is the same as that of DES given we use a similar masking procedure that removes circular regions around bright sources \citepalias{paper1}.} and we therefore follow \citet{Friedrich:2021:CovY3} in assuming the mask can be ignored in the mixed term.

The contribution of the shape noise term is sensitive to the inhomogeneity in the \textit{source-galaxy number density}. Note that this is \textit{not} the inhomogeneity in the image quality/depth. Figure \ref{fig:Survey_homogeneity} illustrates the relevance of this distinction. The left panels show the magnitude limit maps from \decade and from DES Y3, both generated using the \textsc{Decasu} software.\footnote{\url{https://github.com/erykoff/decasu}} There is a clear visual difference in the homogeneity of these maps. However, this does not directly translate into the middle column, which shows the number density map, $n_{\rm gal}$, of the shape catalog. The catalogs in both DES Y3 and \decade are defined with a cut on the $i$-band magnitude, $m_i < 23.5$, which is around the median depth of the \decade survey. This means the long tail to high magnitude limits (Figure \ref{fig:Maglim_dist} below, or also Figure 1 in \citetalias{paper1}) is alleviated through this cut. Note that this cut was placed in DES Y3, and is replicated in \decade, in order to improve the accuracy of our photometric redshift estimates \citep{Myles:2021:DESY3}. The middle column of Figure \ref{fig:Survey_homogeneity} shows that the variation in $n_{\rm gal}$ is in fact similar across \decade and DES Y3. In summary, while the \decade survey's observing conditions are more inhomogeneous than DES Y3, the number density of the resulting source-galaxy catalog is similarly homogeneous to that of DES Y3.

Given our survey mask has a regular geometry and similar small-scale feature compared to DES Y3, we rely on the findings on \citet{Friedrich:2021:CovY3} and do not revalidate the $\mathbf{C}_{\rm mix}$ and $\mathbf{C}_{\rm CV/SSC/cNG}$ terms in Equation \eqref{eqn:cov:contributions}. In this context, we only need to check the impact of inhomogeneity on the shape noise-related terms. As a reminder, the covariance model of \citet{Friedrich:2021:CovY3} included the survey mask in $\mathbf{C}_{\rm SN}$ term, using the method of \cite{Troxel:2018:Cov}, and ignored it in the mixed term (as it contributed only 1\% to the posterior scatter, as mentioned above). The variations in galaxy number density within the mask was not included when accounting for the mask. In Appendix \ref{appx:cov}, we confirm that these variations have a minimal impact in the covariance estimate for the \decade dataset and can therefore be ignored, as was done in DES Y3.

\begin{table}
    \centering
    \begin{tabular}{l l l}
        Parameter & Fiducial & Prior \\
        \hline
        \multicolumn{3}{c}{Cosmology}\\
        \hline
        $\Omega_{\rm m}$ & 0.27 & $\mathcal{U}(0.1,0.9)$ \\
        $\Omega_{b}$ & 0.048 & $\mathcal{U}(0.03,0.07)$\\
        $h$ & 0.69 & $\mathcal{U}(0.55,0.91)$ \\
        $A_s \times10^9$ & -- &  $\mathcal{U}(0.5,5)$ \\
        $\sigma_8$ & 0.846 & --  \\
        $n_s$ & 0.97 &$\mathcal{U}(0.87,1.07)$ \\
        $\Omega_{\nu} h^2$ & 0.00083  & $\mathcal{U}(0.0006, 0.00644)$  \\[5pt]  
        \hline
        \multicolumn{3}{c}{Intrinsic Alignments}\\
        \hline
        $a_{1}$ & 0.19  & $\mathcal{U}(-4,4)$ \\
        $a_{2}$ & -0.47 & $\mathcal{U}(-4,4)$ \\
        $\eta_{1}$ & -2.6 & $\mathcal{U}(-4,4)$\\
        $\eta_{2}$ & 3.38 & $\mathcal{U}(-4,4)$\\
        $b_{\rm ta}$ & 0.0066 & $\mathcal{U}(0,2)$ \\[5pt]  
        \hline
        \multicolumn{3}{c}{Redshift calibration}\\
        \hline
        $\Delta z_1$ & 0.0 &$\mathcal{N}(0, 0.0163)$ \\
        $\Delta z_2$ & 0.0 &$\mathcal{N}(0, 0.0139)$ \\
        $\Delta z_3$ & 0.0 &$\mathcal{N}(0, 0.0101)$ \\
        $\Delta z_4$ & 0.0 &$\mathcal{N}(0, 0.0117)$ \\[5pt]  
        \hline
        \multicolumn{3}{c}{Shear calibration}\\
        \hline
        $m_1$ & 0.0 &$\mathcal{N}(-0.00923, 0.00296)$  \\
        $m_2$ & 0.0 &$\mathcal{N}(-0.01895, 0.00421)$ \\
        $m_3$ & 0.0 &$\mathcal{N}(-0.04004, 0.00428)$ \\
        $m_4$ & 0.0 &$\mathcal{N}(-0.03733, 0.00462)$ \\
        \hline
    \end{tabular}
    \caption{Cosmology and nuisance parameters in the baseline $\Lambda$CDM model. Uniform distributions in the range $[a,b]$ are denoted $\mathcal{U}(a,b)$ and Gaussian distributions with mean $\mu$ and standard deviation $\sigma$ are denoted $\mathcal{N}(\mu,\sigma)$. The column ``Fiducial'' refers to the parameters used for generating the synthetic data vector in all the simulated likelihood tests. The redshift calibration and shear calibration parameters are listed for each of the four tomographic bins. See Section 3.2 of \citetalias{paper2} for the binning scheme.}
    \label{tab:params}
\end{table}

\subsection{Likelihood and sampling}
\label{sec:likelihood}

We fit the model above to the cosmic shear measurements of $\xi_{\pm}$ using a Markov Chain Monte Carlo (MCMC) approach. We assume a Gaussian likelihood $\mathcal{L}$, with
\begin{equation} \label{eq:likelihood} 
\ln \mathcal{L} ( \xi_{\pm,d} | \boldsymbol{p})
\propto -\frac{1}{2}\bigg(\xi_{\pm,d} - \xi_{\pm,m}(\mathbf{p})\bigg)\mathbf{C}^{-1} \bigg(\xi_{\pm,d} - \xi_{\pm,m}(\mathbf{p})\bigg), 
\end{equation} 
where $\xi_{\pm,d}$ and $\xi_{\pm,m}$ are the $\xi_{\pm}$ data vector measured in the data and that predicted by the theoretical model of Equation~\ref{eq:xipm}, respectively; $\textbf{C}^{-1}$ is the inverse covariance described in Section~\ref{sec:cov}; $\textbf{p}$ is a vector of cosmological model parameters and nuisance parameters listed in Table~\ref{tab:params}. The Bayesian posterior is proportional to the product of the likelihood $\mathcal{L}$ and the prior $P$,
\begin{equation} 
P(\mathbf{p}|\xi_{\pm,d}) \propto \mathcal{L} ( \xi_{\pm,d} | \mathbf{p})P(\mathbf{p}). 
\end{equation} 
In Table~\ref{tab:params} we list the priors as well as fiducial values of all the model parameters. 

We perform the parameter inference using the \textsc{CosmoSIS} package \citep{Zuntz2015}. In particular, we use the \texttt{Nautilus} sampler \citep{Lange:2023:Nautilus} for most of our tests, and have cross-checked that the results are consistent with those obtained using the \texttt{Polychord} sampler \citep{Handley:2015:Polychord}. We also produce results using the \textsc{MultiNest} sampler \citep{Feroz:2009:Multinest} for an additional comparison point. The exact parameters we use for the samplers are listed in Table~\ref{tab:sampling}.

\begin{table}
    \centering
    \begin{tabular}{ll l}
      Sampler  & Parameter & Value \\
      \hline
        \texttt{Polychord} & &  \\
& \texttt{fast\_fraction} & 0.1 \\
& \texttt{live\_points} & 500 \\
& \texttt{num\_repeats} & 60 \\
& \texttt{tolerance} & 0.01 \\
& \texttt{boost\_posteriors} & 10.0 \\
\hline
\texttt{Nautilus} & & \\
& \texttt{n\_live} & 1500 \\
& \texttt{discard\_exploration} & True \\
\hline
\texttt{Multinest} & & \\
& \texttt{live\_points} & 500 \\
& \texttt{efficiency} & 0.3 \\
& \texttt{tolerance} & 0.1 \\
& \texttt{constant\_efficiency} & False \\
   \hline
\end{tabular}
    \caption{Parameters used in the different samplers tested in Section~\ref{sec:sampling_s8}. \texttt{Nautilus} is the primary sampler used in this project. \texttt{Polychord}, which is the default sampler in DES analyses, is used to validate our results obtained with  \texttt{Nautilus}. We also use \texttt{Multinest} as an additional comparison point.}
    \label{tab:sampling}
\end{table}
 
\section{Validation of fiducial model}
\label{sec:validation}

In this section we validate our model --- specifically, testing that it is sufficient for obtaining unbiased constraints --- by generating simulated data vectors\footnote{These are generated using the theory predictions from the \textsc{Cosmosis} modelling framework and its associated packages. Thus, they are theory datavectors with no noise.} and analyzing them using the pipeline. We generate a fiducial data vector using the parameters in the ``Fiducial'' column in Table~\ref{tab:params}. In particular, the IA parameters have fiducial values taken to be the best-fit parameters from the DES Y3 3$\times$2pt\footnote{The 3$\times$2pt nomenclature of DES refers to the combination of two-point correlation functions of three probes: galaxy clustering, galaxy galaxy-lensing, and cosmic shear.} results \citep{DES2022}. The cosmological parameters were first chosen to be generic ($\Omega_{\rm m}=0.3$, $\sigma_8=0.8$) in the initial testing, but after unblinding of our final cosmology constrains (see \citetalias{paper4}) we updated the values to more closely match said constraints. Regardless, the tests in this paper are quite insensitive to the assumed cosmology and to the relatively small shifts we applied pre/post-unblinding. We note that the change post-unblinding was only to the simulated datavector analysed in the following tests. None of the modelling decisions were changed post-unblinding.

\begin{figure}
    \centering
    \includegraphics[width=\columnwidth]{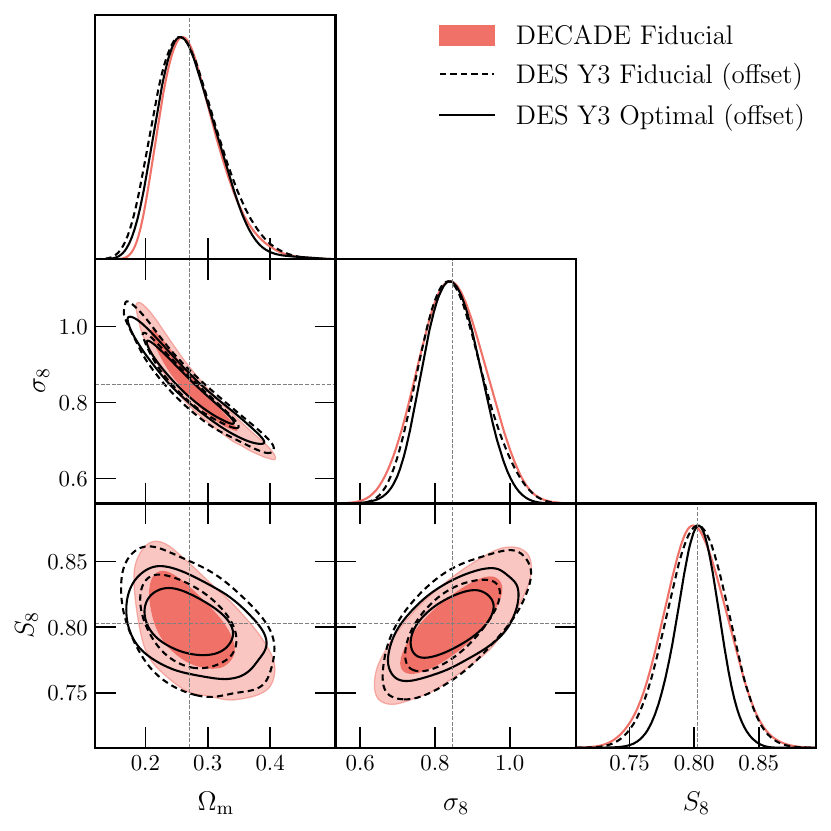}
    \caption{Forecasted constraints on $\Omega_{\rm m}$, $\sigma_8$ and $S_8$ using a simulated data vector corresponding to the \decade dataset (red). The cross-hair indicates the input cosmology, showing that our pipeline adequately recovers the input values. The DES Y3 constraints from \citet{Secco2021} and \citet{Amon2021} are overlaid for comparison. To enhance comparisons of constraining power, we shift the DES posteriors to be centered on the simulated data vector's input cosmology. Both the fiducial and the optimal constraints from DES include information from shear ratios \citep{Sanchez2022}, resulting in tighter constraining power and slightly different degeneracy directions.}
    \label{fig:fiducial}
\end{figure}

In Figure~\ref{fig:fiducial}, we show the resulting cosmological constraint of our fiducial model in the plane spanned by $\Omega_{\rm m}$, $\sigma_8$, and $S_8 \equiv \sigma_8 \sqrt{\Omega_{\rm m}/0.3}$. The input cosmology is indicated by the cross-hair and shows our pipeline recovers the input parameters. This is a simple check, but can be non-trivial due to the impact of parameter projection effects (also called ``prior volume'' effects) --- caused by a high-dimensional parameter space and/or sufficiently asymmetric posteriors --- as in such cases the marginalized 1D constraints are not guaranteed to be centered at/around the truth. Of the parameters shown in Figure~\ref{fig:fiducial}, we focus on the derived parameter $S_8$, which has been widely adopted in the lensing community since it approximates the most constraining direction in the $\sigma_8$-$\Omega_{\rm m}$ plane \citep{Jain1997}. Our simulated analysis shows $S_8=0.801^{+0.024}_{-0.024}$, where the uncertainties denote the 68\% confidence interval computed from the posterior.

We also overlay the DES Y3 constraints \citep{Secco2021, Amon2021} for comparison: $S_8 = 0.759^{+0.025}_{-0.023}$ and $S_8 = 0.772^{+0.018}_{-0.017}$ for the ``Fiducial'' and ''Optimal'' cases, respectively. The ``Optimal'' case cuts fewer scales from the final analysis, and therefore provides better constraints. Our focus is the relative size of the contours, and not the relative locations (as we are doing a simulated analysis for the \decade dataset and have complete freedom in the location of the resulting contour), so we shift the DES posteriors to be centered on the input cosmology of the simulated data vector. The figure shows the \decade constraints are expected to be similar to the DES Y3 ``Fiducial'' constraints, and broader that the DES Y3 ``Optimal'' constraints. Our scale cuts match the former setup. We also note that all DES constraints shown here include information from shear ratios \citep{Sanchez2022}. We perform a more detailed comparison of the \decade and DES Y3 constraints in \citetalias[][]{paper4} and find the \decade result slightly outperform the DES Y3 ``Fiducial'' result once all analysis choices are matched.

In the following subsections we perform a number of stress-tests on the model. We summarize all the tests in Figure~\ref{fig:Sim_Validation}, where each line shows the marginal $1\sigma$ and $2\sigma$ intervals in $S_8$, $\Omega_{\rm m}$ and $\sigma_8$ for a given modeling scenario, and compare it with the intervals of the fiducial constraints (vertical shaded bands) that was also shown in Figure~\ref{fig:fiducial}. First, Section~\ref{sec:scale_s8} confirms that our fiducial scale cuts (derived in Section~\ref{sec:scalecuts}) alleviate biases from baryonic imprints in the data vector. Next, Section~\ref{sec:model_s8} discusses shifts in constraints when assuming different models for the nonlinear matter power spectrum and the IA. In Section~\ref{sec:nuisance_s8} we examine the sensitivity of our constraints to the priors on the nuisance parameters, and in Section~\ref{sec:sampling_s8}, the sensitivity of the constraints to the choice of sampling methods.

In Appendix \ref{appx:Contam2Cosmo}, we also verify that (1) the contamination from correlations of the point-spread function has negligible impact on our cosmology constraints, and; (2) there is no statistically significant detection of $B$-modes in all auto- and cross-correlations of our data vector. Both tests complement those found in \citetalias{paper1}.

\subsection{Robustness to small-scale modeling / baryons}
\label{sec:scale_s8}

\begin{figure*}
    \centering
    \includegraphics[width=1.8 \columnwidth]{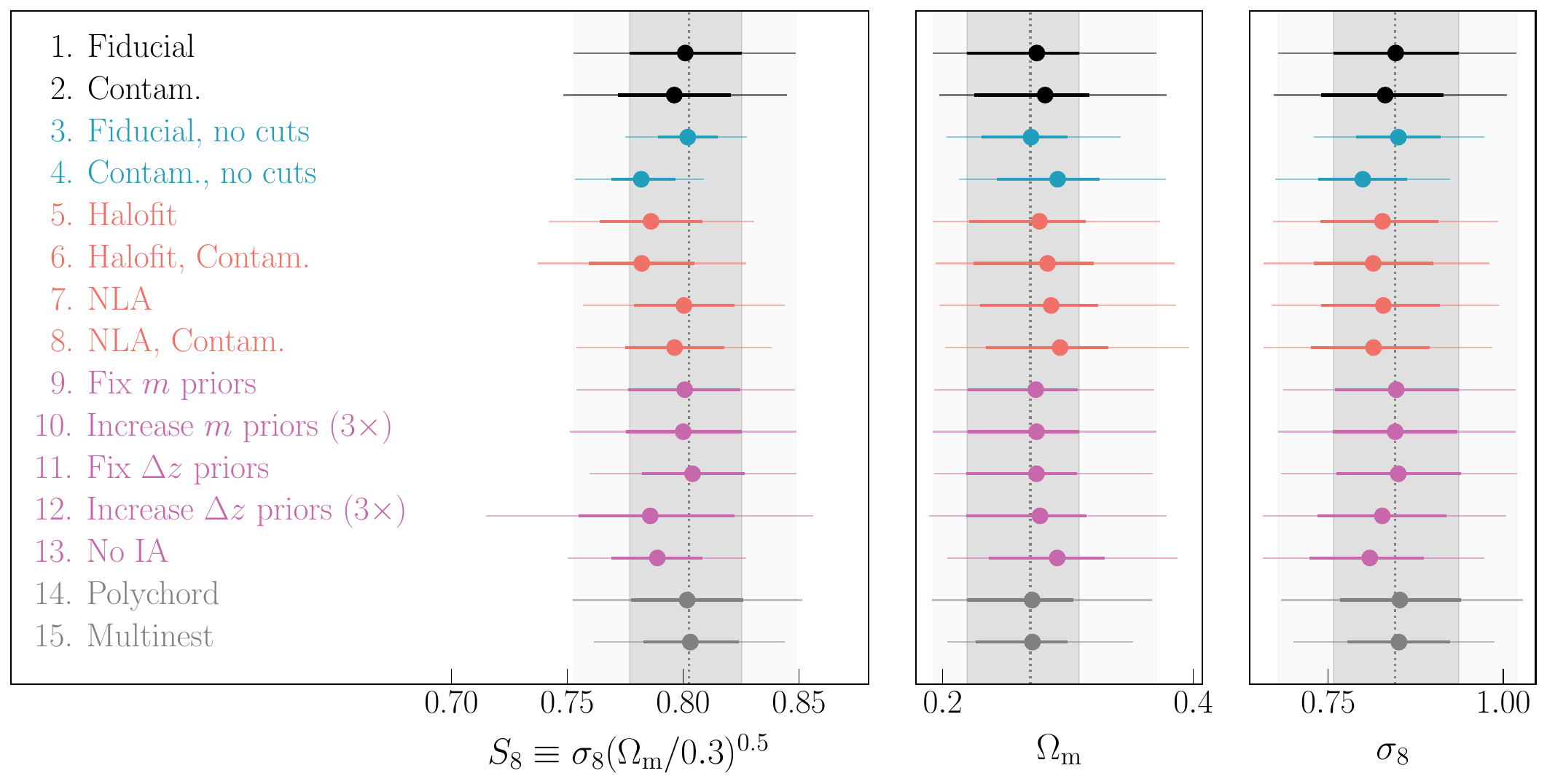}
    \caption{Constraints on $S_8$, $\Omega_{\rm m}$ and $\sigma_8$ from simulated data vectors, under different analysis choices. This series of tests (detailed in Section~\ref{sec:validation}) ensures that the modeling choice we adopted for the data is robust, including the mitigation of small-scale baryonic effects, the assumptions of the nonlinear power spectrum and intrinsic alignment, the priors on the nuisance parameters, and the choice of sampling approaches in the inference process. In this plot, as well as later Figures~\ref{fig:SurveySplits} and ~\ref{fig:SurveySplitsProperties}, the constraints shown are the marginalized mean, 1$\sigma$ and 2$\sigma$ uncertainties. The vertical shaded region correspond to the 1$\sigma$ (dark) and 2$\sigma$ (light) intervals of the fiducial case. Here, we also show the true input value to the simulated data vectors with the vertical dotted line. These are $\Om = 0.27, \sigma_8 = 0.846, S_8 = 0.80$. Note that line 1 (``Fiducial'') corresponds to the marginal constraints shown in Figure~\ref{fig:fiducial}.}
    \label{fig:Sim_Validation}
\end{figure*}

As discussed in Section~\ref{sec:scalecuts}, we remove measurements from small-scale correlations during our inference as there are large uncertainties in the matter power spectrum model on such scales, primarily due to the effects of baryons \citep{Chisari2018BaryonsPk}. This section tests the impact of small-scale baryonic physics on cosmological constraints, and the effectiveness of scale-cuts in mitigating their impact In Figure~\ref{fig:Sim_Validation}, we show constraints with and without scale cuts, for both the fiducial data vector (lines 1 and 3) and also one contaminated by baryon effects (lines 2 and 4). The contaminated data-vector is the same one used to determine the scale cuts described in Section~\ref{sec:scalecuts}. The case ``without scale cuts'' uses all measured data points from $2.5 \arcmin < \theta < 250 \arcmin$ for the data-vector in question.

We first compare line 1 and 3. Without baryonic contaminations, we see that both constraints are centered on the input values of the cosmological parameters, and the scale cuts degrade the constraints by a factor of $\approx$2. This highlights the significant cosmological information in the scales that we currently remove due to baryonic effects. Next, we compare the pairs of line 3 and 4, and then line 1 and 2. If we do not use scale cuts, the contamination from baryonic effects shifts the extracted constraints. We evaluate the shift in the $\Omega_{\rm m}$-$S_8$ plane to be 0.67$\sigma$. Once we apply the scale cuts, the shifts reduce to 0.01$\sigma$, which is within our criteria of 0.3$\sigma$. Note that the shift moves in the direction that decreases $S_8$, which is expected as baryonic feedback suppresses the matter power spectrum on the small scales relevant for weak lensing \citep{Chisari2018BaryonsPk, Amon2022S8Baryons}.

\subsection{Robustness to nonlinear power spectrum and IA models}
\label{sec:model_s8}

We have made certain choices in the modeling pipeline based on our best, current knowledge (which is informed by the latest cosmic shear analyses, \eg \citet{Secco2021, Amon2021, Asgari2021, DESKiDS2023, Li2023}) on the validity and expected accuracy of each model. In particular, the nonlinear power spectrum and the IA model are chosen based on arguments detailed in Section~\ref{sec:model}. In this section we check the shift in $S_8$ constraints if under alternative choices for these models. Note that in all cases, the changes are made only to the analysis pipeline; the simulated data vector is still fixed to our fiducial setup.

Line 5 in Figure~\ref{fig:Sim_Validation} shows the $S_8$ constraints when using \textsc{HaloFit} as our model for the nonlinear matter power spectrum. Using \textsc{HaloFit} to analyze a \textsc{HMCode}-generated data vector shifts the $S_8$ constraint to a slightly lower value, which is consistent with what is seen in \eg\citet{Secco2021}. The shift is because at fixed cosmology, the \textsc{HaloFit} model predicts slightly more power on nonlinear scales relative to \textsc{HMCode} \citep{Mead2016, Mead2020a}. For the IA model comparison, line 7 in Figure~\ref{fig:Sim_Validation} shows that the use of NLA (with two free parameters) tightens the cosmological constraints slightly ($\approx \!8\%$) relative to TATT, which has five free parameters. TATT is a more general model for intrinsic alignments and includes three additional parameters (see Table~\ref{tab:params}). The reduced flexibility of the NLA model (relative to that of TATT) restricts the parameter space and tightens the cosmology constraints.

We also verify that our scale cuts, determined in Section~\ref{sec:scale_s8}, are effective at minimizing baryon-driven biases in $S_8$ even if we use alternative choices in the analysis pipeline. If we used \textsc{HaloFit} as our nonlinear matter power spectrum model, we find a baryon-driven shift of 0.08 $\sigma$ in $S_8$ (comparing line 5 and 6 in Figure~\ref{fig:Sim_Validation}). Similarly, if we change the IA model to NLA we find a shift of only 0.05$\sigma$ in $S_8$ (comparing line 7 and 8 in Figure~\ref{fig:Sim_Validation}).

\subsection{Impact of priors on nuisance parameters}
\label{sec:nuisance_s8}

We next check the impact of the different nuisance parameter marginalization on the final constraining power. Starting from the fiducial analysis setup, we either fix or broaden priors for the shear calibration $m$ and the mean redshift calibration $\Delta z$, relative to the fiducial setup (Table \ref{tab:params}). We also run a variant where we fix the IA parameters. Lines 9-13 of Figure~\ref{fig:Sim_Validation} shows the different constraints from changing the priors on these nuisance parameters. The shear calibration uncertainties ($m$ priors) are a subdominant contribution to the final cosmology constraints, as narrowing/widening these parameter priors has negligible impact on the final constraining power (line 9 and 10). The parameters associated with the redshift uncertainty show some small effect as fixing their values reduces the uncertainty in $S_8$ by $\approx\! 8\%$ (line 11) and widening the prior by three times increases the uncertainty in $S_8$ by $\approx\! 20\%$ (line 12).\footnote{Figure \ref{fig:Sim_Validation} shows that broadening the redshift calibration uncertainties, also shifts the mean of the $S_8$ marginal posterior away from the input value by around $0.5\sigma$ of the posterior width. This is a manifestation of projection effects, which can occur as the prior volume is expanded. We note that such projection effects are not present in our fiducial analysis (Line 1), where the redshift calibration uncertainties are \textit{not} artificially broadened.} 
Finally, fixing all IA parameters (Line 13) improves the constraints by 23\% (13\%) relative to our fiducial (NLA) case. These results indicate that, after applying scale cuts, the primary limitation on the constraining power of cosmology parameters is the uncertainties in the intrinsic alignment model --- and not from the shear and redshift calibrations. The latter limitation from IA uncertainties is also explored in more detail in the data analysis of \citetalias{paper4}.

\subsection{Robustness to sampling choices and noise}
\label{sec:sampling_s8}

As discussed in Section~\ref{sec:likelihood}, all cosmology results for our tests (including those found below and those in \citetalias{paper1}) are obtained using the \texttt{Nautilus} \citep{Lange:2023:Nautilus} sampler, whereas our main cosmology results will be extracted using the \texttt{Polychord} sampler \citep{Handley:2015:Polychord} run on the high accuracy setting (Table \ref{tab:sampling}). We show in line 1 and 14 of Figure~\ref{fig:Sim_Validation} that under the fiducial settings, the results from \texttt{Nautilus} and \texttt{Polychord} are very consistent with each other both in the mean and the width of the posterior on $S_8$. The means are shifted by less than $0.05\sigma$ of each other and the posterior widths are within $4\%$ of each other. We also show in line 15 the same analysis run with the \texttt{Multinest} sampler, which show slightly underestimated error bars (by about 7\% in $S_8$). This has been pointed out in previous work \citep{Lemos2023} and we confirm that result here.

In summary, the \texttt{Nautilus} sampler is consistent with constraints from \texttt{Polychord}, where the latter is a widely used sampler in DES analyses and other works. We therefore adopt \texttt{Nautilus} for all remaining constraints presented in this work and for those presented in \citetalias{paper4}. 

\section{Impact of spatial inhomogeneities}\label{sec:SpatialInhomog}

We now quantify the impact of any spatially varying effects on our cosmology constraints, using a fully data-driven approach. As discussed previously, the \decade cosmic shear catalog is constructed using image data from a wide variety of community-led campaigns, which in turn results in a variety of observing conditions and image quality. This is in contrast to other dedicated lensing surveys --- such as DES, KiDS, and Subaru Hyper Suprime Cam \citep[HSC,][]{Miyazaki:2018:HSC} --- where homogeneity in image quality and depth is a target of the survey observing strategy \citep[\eg][]{Neilsen:2019:ObsStrat, Kuijken:2019:ObsStrat, Aihara:2022:ObsStrat}.

We reiterate a subtlety discussed above in Section \ref{sec:cov}: the impact of survey inhomogeneity on source galaxy number counts, $n_{\rm gal}$, is largely alleviated through our sample selection (a cut on apparent magnitude $m_i < 23.5$). Figure \ref{fig:Survey_homogeneity}, and its associated discussion in Section \ref{sec:cov}, illustrates this point. However, this \textit{does not} mean the \decade survey is ``as homogeneous'' as DES Y3. The observing conditions of the survey --- which we will describe below in Section \ref{sec:SpatialInhomog:splits} --- will still impact the distribution of measured galaxy properties in a spatially dependent manner. For example, regions of the sky with deeper imaging (higher magnitude depth) will contain galaxies whose properties are measured to better precision. Such an effect can still impact the measurements of shear, redshifts, \textit{etc.}, and therefore, the final cosmology constraints.

We have so far performed a number of tests that can probe the impact of inhomogeneity. For example, in \citetalias{paper1}, we performed a series of null tests on the shear catalog and showed there are no residual systematics at the precision level of the catalog. In \citetalias{paper2}, we showed that two independent methods of photometric redshift estimation --- which will have different responses to inhomogeneous survey conditions --- result in consistent $n(z)$ estimates. These are only implicit tests of the impact of inhomogeneity. In this paper, we perform an explicit, end-to-end test for the impact of inhomogeneity on our cosmology constraints. We describe below our methodology for this test.

\subsection{Methodology}\label{sec:SpatialInhomog:methods}

\begin{figure*}
    \centering
    \includegraphics[width=1.8\columnwidth]{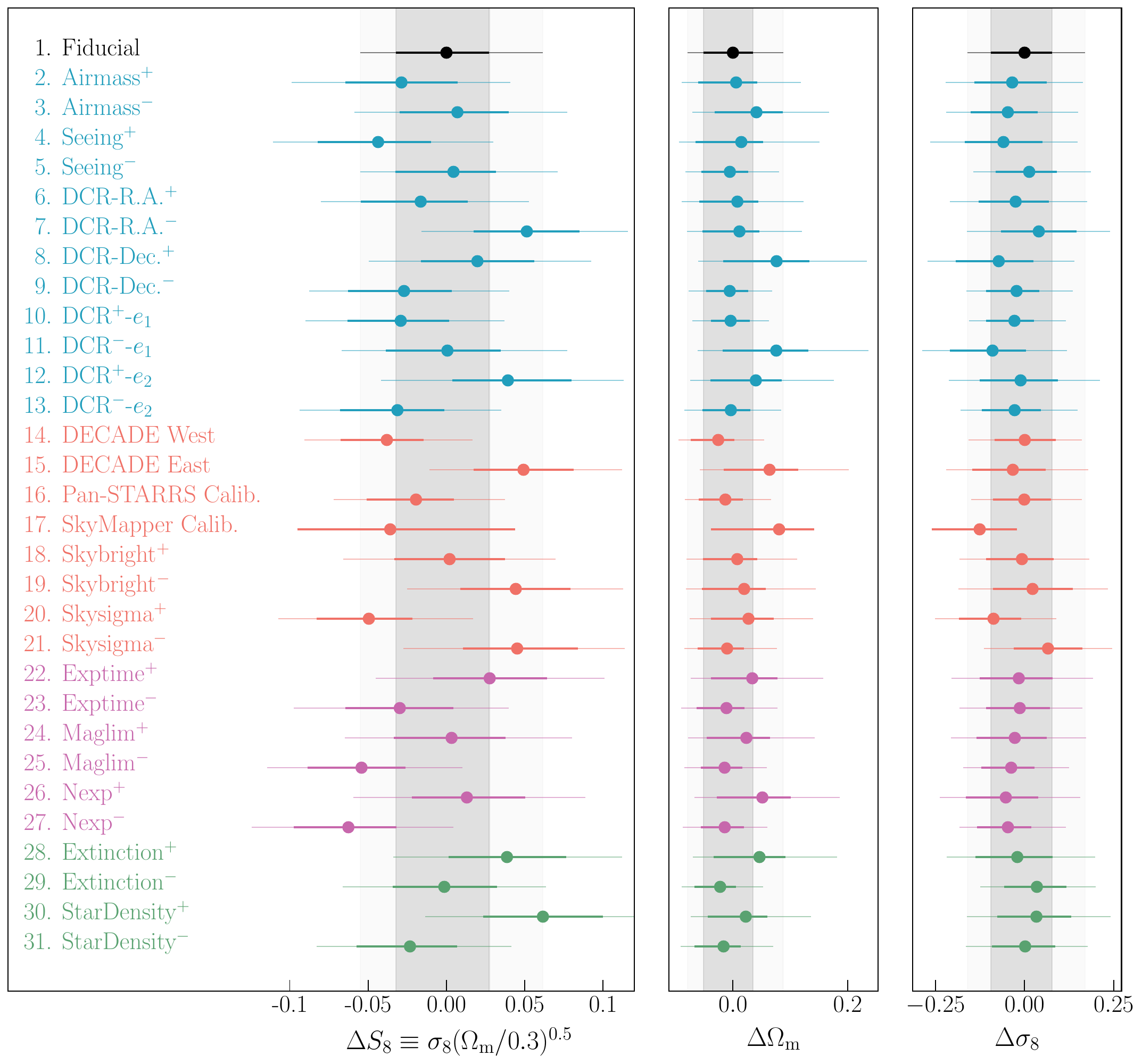}
    \caption{The constraints on $S_8$ from different subsets of the \decade footprint, while consistently rederiving all calibrations, redshift distributions, covariance matrices, \textit{etc}. The first, ``Fiducial'' constraint is using data the entire dataset. The rest, from top to bottom, split the survey based on: airmass, seeing (PSF width), differential chromatic refraction (DCR) effects in the R.A., Dec., and $e_{1,2}$ coordinates, the east/west regions, the $\rm Dec. \gtrless -30$ region where photometric calibrations used Pan-STARRS/SkyMapper, the sky background brightness and variations, exposure time, magnitude limit, number of exposures, interstellar extinction coefficients, and \textit{Gaia} stellar number density. Analyses with $X^+$ ($X^-$) use the area with higher (lower) values than the median. The data vectors in each subset have essentially independent noise realizations and so the cosmology constraints can exhibit relative shifts of up to $3\sigma$ due to just statistical fluctuations. We show the parameter constraints relative to the fiducial estimate, with thick and thin horizontal lines denoting the $68\%$ and $95\%$ intervals for each constraint, respectively. For visibility reasons, we do not show the $95\%$ interval for the ``SkyMapper'' subset alone. Most (all) deviations are within $1\sigma$ ($2\sigma$) of the fiducial results. The constraints from a subset and its complement are always within $2\sigma$ as well. No split exhibits shifts in cosmology that are larger than the expected variation from shape noise and cosmic variance. The footprints corresponding to the splits are shown in Figure \ref{fig:SplitsMaps}.}
    \label{fig:SurveySplits}
\end{figure*}

Our end-to-end test involves splitting the source-galaxy sample into subsets (usually halves) based on some criteria (see Section \ref{sec:SpatialInhomog:splits} and \ref{sec:SpatialInhomog:Objsplits}) and redoing all measurement and calibration steps; this includes the shear calibration (\citetalias{paper1}), redshift estimation/calibration (\citetalias{paper2}), analytical covariance generation, data vector measurements, and cosmological inference (this work). We consider splits both on the survey area and on the galaxy properties. The former is a spatial split defined using the inhomogeneity in the survey image quality, whereas the latter can isolate subsets of the catalog that may be more susceptible to systematics of a specific kind. We briefly detail all the steps in the pipeline that are rerun for each subset:

\textit{Shear weights and response grid.} \citetalias{paper1} and \citetalias{paper2} detail the weights used in the different pieces of the cosmology analysis pipeline. In particular, we use shear weights (inverse-variance weights) and responses, both defined on grids of signal-to-noise and size. In our inhomogeneity tests here, these two grids are regenerated for the exact samples defined in each split. While it is fine to use the fiducial weights with no changes (this only risks performing a sub-optimal analysis but not a biased one) it is important to define the response grid consistently, as the object response is needed to make a robust estimate of the $n(z)$ for a given sample \citep[][\citetalias{paper2}]{Macrann2022ImSim}. For this reason, we remake the weight and response grids for each subset.

\textit{Shear Calibration.} As described above in Section \ref{sec:model:Calib}, all cosmology analyses of weak lensing require some calibration of the multiplicative bias $m$. This is often done using image simulations of the survey, which is also the approach in \decade as detailed in \citetalias{paper1}. We use the existing, simulated catalog and repeat the estimates of $m$, while incorporating the new selection function (which defines the split) that is being applied to the data.

\textit{Redshift estimation.} For our cosmology analysis, we estimate the $n(z)$ using self-organizing maps \citep[SOMs;][]{Kohonen:1982:SOM, Kohonen:2001:SOM, Carrasco:2014:SOM, Masters:2015:SOM} and the technique is denoted as SOMPZ \citep{Buchs2019, Myles:2021:DESY3}. One aspect of this method is classifying the source galaxies into a set of phenotypes, $\hat{c}$ (see \citetalias{paper2}). For these split tests, we do not reassign galaxies to a given phenotype and simply use the existing classification.\footnote{This choice induces no bias in the derived $n(z)$ and only degrades their potential ``optimality''; namely the overlap between the redshift distributions of two tomographic bins. We do not quantify this effect in our work.} However, we redo the selection of galaxies into four tomographic bins, and the estimation of the $n(z)$ per tomographic bin. This includes re-estimating the transfer function --- through the \textsc{Balrog} synthetic source catalog built for \decade \citepalias{paper2} --- to connect the true properties of a galaxy with its observed properties. In all cases, we apply the same area/property selections to the galaxy samples. We generate a large number of $n(z)$ realizations, as is done in \citetalias{paper2}, and characterize the uncertainty in the mean using the same method as described in that work. We do not repeat any measurements of the clustering redshifts; these measurements are only used a cross-check in \citetalias{paper2} and do not inform the fiducial $n(z)$ used in our analysis.

\textit{Covariance matrix.} We estimate new $n_{\rm eff}$ and $\sigma_e$ quantities (effective number density and shape noise of source galaxies, respectively) for the sample in our subset. We also recompute the geometry of the survey footprint, incorporating our new selections, and use it in conjunction with the new $n(z)$ to generate our analytic covariance matrix. This is done through \textsc{CosmoCov}, as is described above in Section \ref{sec:cov}.

\textit{Data vector and cosmology.} We remeasure the data vector using only galaxies in the chosen part of the footprint, while using the new tomographic bin assignments and the new shear weights. During our internal tests, the data vectors from this step were all blinded using the same random seed used on the data \citepalias{paper4}. However, our results presented below are the constraints re-evaluated after unblinding the data vector.

\textit{Consistency.} Finally, we must define a metric that allows us to determine whether constraints from different subsets are consistent with each other and with the fiducial constraints. We do so using the standard score (or ``$z$-score'') for $S_8$,
\begin{equation}
    \text{number of }\sigma = \frac{[S_8]_1 - [S_8]_2}{\sqrt{\sigma([S_8]_1)^2+\sigma([S_8]_2)^2}},
    \label{eq:s8_dist}
\end{equation}
where $[S_8]_i$ and $\sigma([S_8]_i)$ are the mean and standard deviation of the posterior for constraint $i$, respectively. Unlike the previous tests, which involved consistency between analysis choices operating on the exact same data vector, this test uses data vectors with different noise realizations that are essentially uncorrelated,\footnote{The shape-noise realizations are completely uncorrelated between two subsets that use different galaxies in their measurements. However, the cosmic shear measurements can still be correlated due to large-scale (cosmological) density fluctuations that are shared by subsets spanning adjacent regions of the sky. We note that shape noise is the dominant contributor to the measurement covariance.} since they access different subsets of galaxies from different subregions of the \decade footprint. For this reason, it is possible for cosmology constraints from a given split to be different purely due to random fluctuations (where the fluctuations are dominated by shape noise, but will also have some cosmic variance component). Thus, our null test is satisfied if the $S_8$ constraints from a given subset are within $3\sigma$ of the constraints from the complementary subset. Though, in practice, we find all our consistency estimates are comfortably below this threshold. Here, we define complementary subset as ``all galaxies not in the subset''. In practice, there can be mild correlations between a subset and its complement. We have quantified these in two limiting cases and verified the standard score estimate, using Equation \eqref{eq:s8_dist}, is changed by 10-15\% at most. See Appendix \ref{appx:SurveySplits} for more details.

In summary, we redefine the sample selection function to include a selection on survey area/object properties, and repeat all measurements that depend on the selection function. We then check if the cosmology constraints made using subsets of our data are statistically consistent with their complement, and also statistically consistent with the fiducial constraint corresponding to the entire \decade dataset.

\subsection{Survey property splits \& results}\label{sec:SpatialInhomog:splits}

\begin{figure}
    \centering
    \includegraphics[width=\columnwidth]{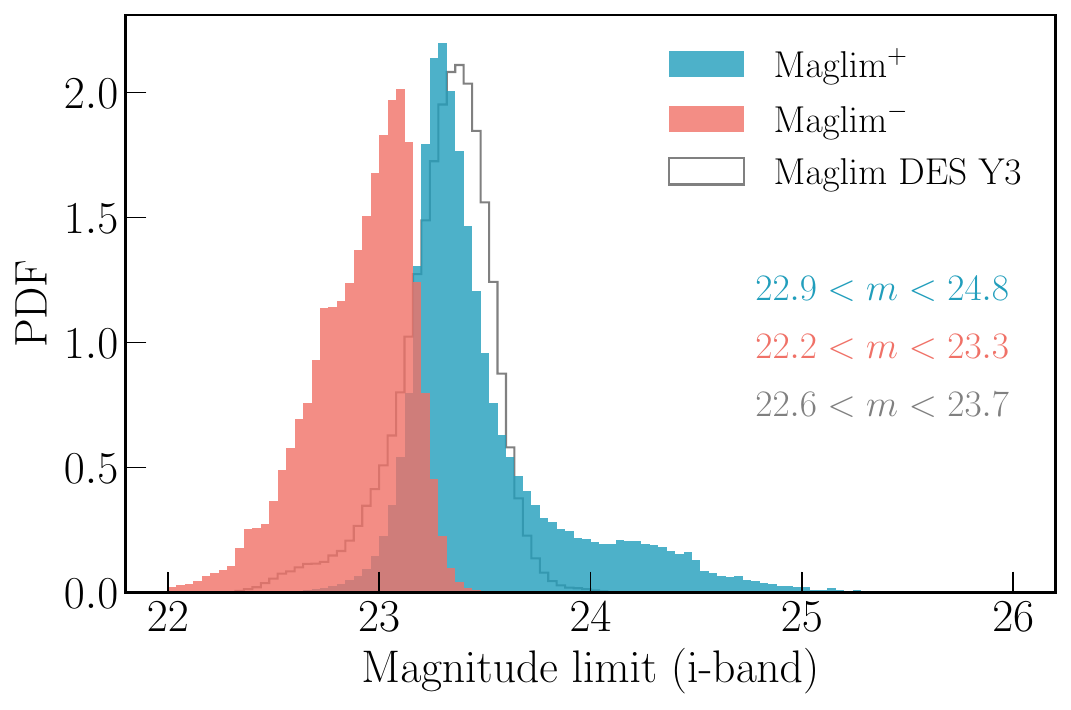}
    \caption{The distribution of magnitude limits in the $i$-band (from a \textsc{HealPix} map of $\texttt{NSIDE} = 4096$) for the two subsets of the \decade footprint split by magnitude limit, and for the DES Y3 footprint. The ranges in the figure denote the $1\%$ and $99\%$ values of the distribution. The Maglim$^+$ region (blue) has twice the width --- defined as the difference between the 99\% and 1\% intervals shown in the Figure --- compared to those of the other two samples. There is a slight, minimal overlap in the two \decade distributions as we plot values from the $\texttt{NSIDE} = 4096$ maps, whereas the selection mask is defined using the downgraded $\texttt{NSIDE} = 1024$ maps (see Section \ref{sec:SpatialInhomog:splits} for details). The overlap has no impact on our qualitative and quantitative discussions.}
    \label{fig:Maglim_dist}
\end{figure}

\begin{figure*}
    \centering
    \includegraphics[width=1.8\columnwidth]{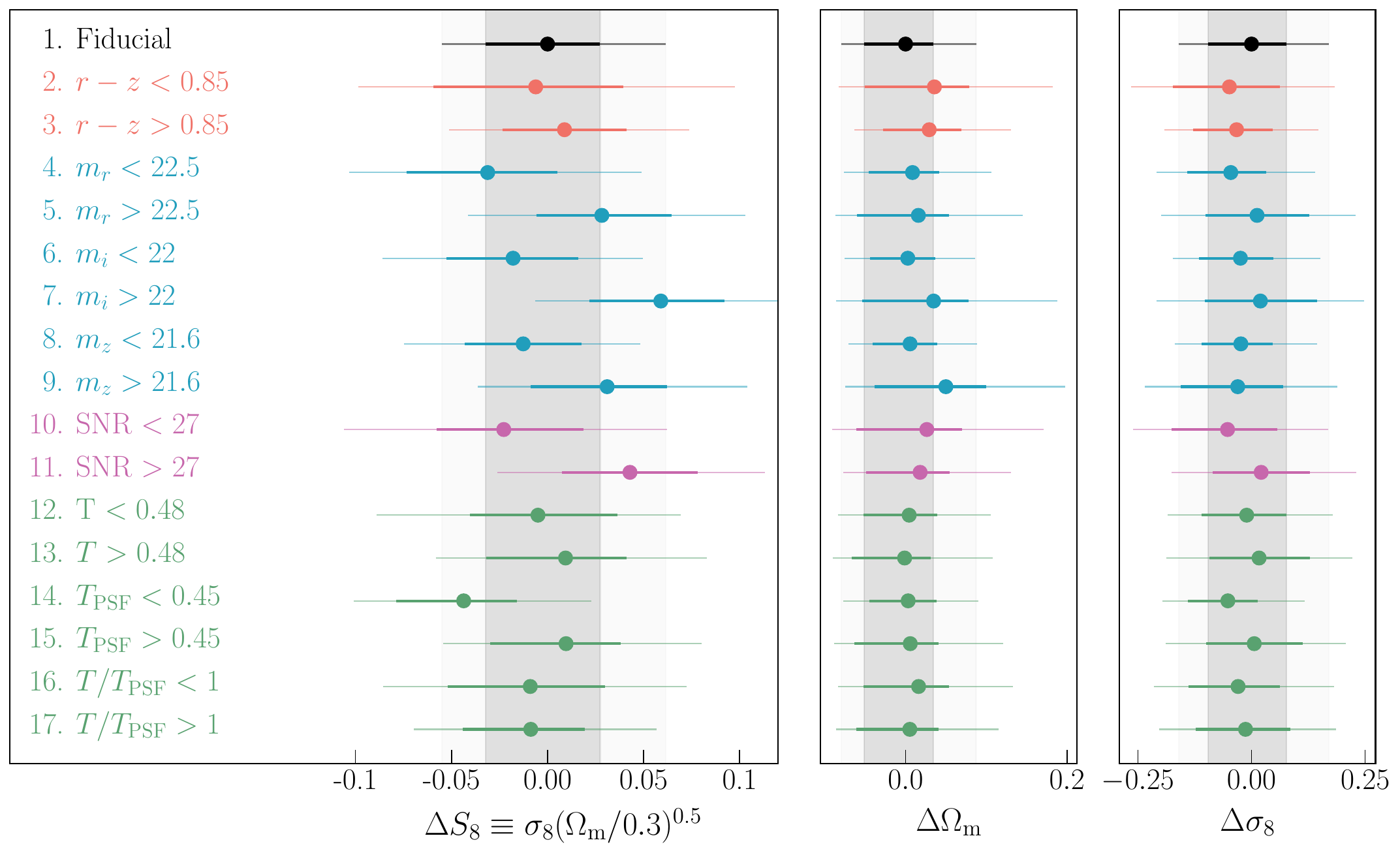}
    \caption{Similar to Figure \ref{fig:SurveySplits} but now splitting the shear catalog on a variety of galaxy properties. From top to bottom, we split the catalog based on: $r - z$ color, magnitude in $riz$ bands, signal-to-noise, the object size (in arcsec$^2$), the PSF size (in arcsec$^2$), and the ratio of object size to PSF size. All quantities are taken from the \textsc{Metacalibration} catalog.}
    \label{fig:SurveySplitsProperties}
\end{figure*}

We first split the sample on a number of different area-dependent quantities, most of which are informed by the survey observing conditions. In \citetalias{paper1} (see their Figure 13), we check that the measured galaxy shapes are not correlated with a wide variety of survey properties: the seeing (PSF width), airmass, differential chromatic refraction (DCR) in $e_1$ and $e_2$ directions, DCR in the right ascension (R.A.) and declination (Dec.) directions, the magnitude limit, exposure time, number of exposures, the standard deviation of the sky background, and the sky brightness. We use this same set of maps in our test here, and also add a number of additional splits defined below. Every survey property map has four variants, one for each of the $griz$ photometric bands. The $g$-band data is not used in any of our lensing analysis (see \citetalias{paper1} for more details). Of the remaining three bands, we choose the $i$-band maps for all our inhomogeneity tests, but note than the $r$ and $z$ bands would be equally viable choices.

We also split the footprint based on the \textit{Gaia} stellar density \citep{Gaia:2021:eDR3} and interstellar extinction maps from \citet{Schlegel:1998:Dust}, which were used in \citetalias{paper1} to define our foreground mask (see discussion in Section 2.1). In all these cases, we split the footprint by sampling the map value for every object in our source-galaxy sample, finding the median value for the sample, and splitting the area of the sky based on that median. This ensures the splits contain an equal number of galaxies. The maps are downsampled to $\texttt{NSIDE} = 4096 \rightarrow 1024$ to improve the overall contiguousness of the selection mask by reducing the presence of disconnected patches in the survey mask. The remaining two split definitions are purely geometric (selections on $\rm R.A.$ and $\rm Dec.$ alone). First, we split our survey on ${\rm Dec.} \gtrless -30$ as all CCD images in regions north (south) of this line used the Pan-STARRS-1 (SkyMapper) measurements --- as found in the ATLAS RefCat2 reference catalog \citep{Tonry:2018:AtlasRefcat2} --- for their photometric calibrations \citep[see][for more details]{Tan:2024}. Second, we split the footprint into an eastern and western footprint, split on ${\rm R.A.} \gtrless 180$.\footnote{We assign the entire SkyMapper-calibrated region (${\rm DEC.} < -30$) to the western footprint as both regions were part of the same, internal image processing run.} Altogether, we define fifteen different ways of splitting the footprint, with each relying on a different survey observing condition.

Figure \ref{fig:SurveySplits} shows the results from all fifteen criteria, which corresponds to thirty different subsets of our footprint. The splits across survey properties often fill our full footprint in a somewhat patchy manner (see Figure \ref{fig:SplitsMaps}), whereas the splits on geometry are distinct, internally contiguous footprints. In all cases, the data in the subsets will have essentially independent realizations of the shape noise (Appendix \ref{appx:SurveySplits}). The results in Figure \ref{fig:SurveySplits} show that for any chosen subset, the constraints from the subset and its complement are within $2\sigma$ of each other. Here, we have used the metric defined in Equation \eqref{eq:s8_dist}. In summary, we find that all thirty splits are within $2\sigma$ of the fiducial result, and each subset is within $2\sigma$ of its complement, which passes our $3\sigma$ criteria. 

\textbf{Area-dependent $\boldsymbol{n(z)}$:} While the above tests do not explicitly probe the impact of spatially correlated variations on the $n(z)$ estimates, they still implicitly check this effect. In particular, the regions of the sky with a higher magnitude limit are also regions with a significantly larger variation in the magnitude limit. If the spatial variation of the magnitude limit has an impact on the true $n(z)$ of our source galaxy sample, then the spatial variation of the $n(z)$ will have a larger amplitude if the variation in the magnitude limit is increased. Figure \ref{fig:Maglim_dist} shows distributions of the magnitude limits under the two subsets, and lists the 1\% and 99\% values of the distributions. These values show that the Maglim$^+$ distribution is nearly twice as a wide as the Maglim$^-$ and DES Y3 distributions. Thus, any spatially varying $n(z)$ effects will be even more prominent in the Maglim$^+$ region. The consistency of the cosmology from the Maglim$^+$ and Maglim$^-$ regions (the constraints are within $1.2\sigma$ as shown in Figure \ref{fig:SurveySplits}) suggests our final cosmology result is not significantly impacted by such area-dependent redshift effects.

\subsection{Object property splits \& results}\label{sec:SpatialInhomog:Objsplits}

We now repeat the analysis of Section \ref{sec:SpatialInhomog:splits} but splitting the sample on \textit{galaxy properties} rather than through a survey mask. While this is not a strict test of spatial inhomogeneities, the property splits will isolate subsets of the catalog that may be more susceptible to inhomogeneity in various observing conditions; for example, the shapes and photometry of galaxies with small sizes can be more strongly affected by PSF variations. All properties used in the splits are taken from the \textsc{MetaCalibration} catalog of \citetalias{paper1}, as this allows any selection-based bias in the shear to be robustly calibrated through the \textsc{MetaCalibration} formalism \citep[\eg][]{Sheldon:2017:Shear}. We split on colors, magnitudes, signal-to-noise ratios, and sizes. The color split is motivated by \citet{Samuroff:2019:IA} and \citet{McCullough:2024:BlueShear} who split the DES Y1 and Y3 sample, respectively, into red and blue galaxies and redo the cosmic shear analysis for each split. Their color split (see Table 1 in \citet{McCullough:2024:BlueShear}) is a $r - z$ cut that \textit{depends on the tomographic bin}. Given the setup of our pipeline, we only perform a non-tomographic $r - z$ split across the full catalog. We choose $r - z \gtrless 0.85$ as it is the median value, across the four bins, used in \citet{McCullough:2024:BlueShear}. We then consider a range of magnitude cuts and size cuts, which probe depth-based and PSF-based effects, respectively. The signal-to-noise cut is correlated with both magnitude and size-based cuts. All the non-color selections were defined using the median value of their distributions in the fiducial shape catalog and the chosen values are listed alongside our results.

Figure \ref{fig:SurveySplitsProperties} shows the constraints from the different splits. Similar to the test in Section \ref{sec:SpatialInhomog:splits}, most (all) constraints from the subsets are within $1\sigma$ ($2\sigma$) of the ``Fiducial'' estimate. Each subset and its complement, for all splits, are also consistent within $2\sigma$ of each other.  These results showcase the robustness of our cosmology results against systematics that preferentially affect certain types of objects in the catalog (\eg objects with small angular sizes) and further highlight that the cosmic shear analysis pipeline can handle data of a less pristine nature. 

Other results associated with the above tests, such as the selection mask definition and the estimated redshift distributions, are presented in Appendix~\ref{appx:SurveySplits}.

These split tests are a powerful method for inferring the impact of systematics on cosmology through a \textit{completely data-driven approach}. Throughout this analysis we did not have to produce any simulations assuming a given systematic effect (and therefore did not need an a-priori understanding of how a given systematic effect impacts the data). The disadvantage of this method is its relative insensitivity --- it is best at indicating whether a systematic is causing a $>3\sigma$ parameter shift. While this is still valuable for catching hidden or poorly understood systematics, it is not ideal for characterizing systematics that have smaller effects on the posterior constraints. One obvious extension to our approach is to link together the parameter shifts from different splits and determine consistency in cosmology constraints in an ensemble sense, i.e. by summarizing the distribution of shifts (across different splits) through a $p$-value metric. This requires a precise, robust covariance matrix for the correlation function measurements of the different subsets. One approach to do so is using simulation-based estimates. We have not pursued this given the computational cost. However, the calculation is tractable and so could be pursued in the future.

\section{Summary}
\label{sec:summary}

In this paper, we outline the methodology used to perform the cosmological inference of the Dark Energy Camera All Data Everywhere (DECADE) shear catalog. Our modeling framework and implementation builds heavily on the DES Year 3 cosmic shear analysis \citep{Secco2021, Amon2021}, and also incorporates recent updates \citep{DESKiDS2023}. After describing the basic modeling choices, we perform a series of specific tests to: (1) ensure that our baseline analysis choices are robust to different choices in the model, and; (2) stress-test our pipeline by applying the same analysis pipeline on forty-six different subsets of the dataset, each defined as a selection on observing conditions or galaxy properties. Some of these tests are particularly important as the \decade dataset is a compilation of very different DECam observing campaigns and thus, the characteristics of this dataset are more inhomogeneous than conventional weak lensing surveys. 

We summarize our main points below:
\begin{itemize}
    \item Our fiducial model uses \textsc{HMCode} for the nonlinear matter power spectrum but does not marginalize over baryonic effects. Instead we remove the small-scale data points that can be contaminated from the presence of such baryonic effects. Our fiducial analysis uses similar scale cuts to the Fiducial DES Y3 analysis. We use the TATT intrinsic alignment model and a one-parameter shift (per tomographic bin) for the redshift uncertainty. 
    \item For the covariance matrix, we use \textsc{CosmoCov} to generate an analytical covariance matrix that includes both Gaussian and non-Gaussian contributions. We also specifically test the shape noise model for the covariance matrix using simulations, in Appendix \ref{appx:cov}, and find that the inhomogeneity in the galaxy distribution (at the level seen in \decade) does not appreciably impact the model predictions.
    \item Our expected constraint power is similar to the DES Y3 ``Fiducial'' analysis constraints but 30\% worse than the DES Y3 ``Optimal'' analysis constraints, given the latter removes significantly fewer scales in its analysis (Section \ref{sec:validation}).
    \item We stress-test our methodology by rerunning the end-to-end pipeline (rom catalog to cosmology) after splitting the \decade catalog into subsets based on various observing conditions or measured galaxy properties. Overall, we tested twenty-three different definitions of the split (forty-six subsets in total). These end-to-end tests check whether the spatially inhomogeneous nature of the subset can introduce significant biases (relative to the posterior width) in the cosmological constraints. Over the forty-six runs, the resulting cosmological constraints are mostly (always) within 1$\sigma$ (2$\sigma$) of the fiducial constraints, indicating all subsets of the data show statistically consistent cosmological constraints. 
    \item This is the first time the above end-to-end test is done with such completeness in a weak lensing survey. This was made possible by self-consistently recalibrating the shear estimates, remeasuring the redshift distributions and calibrations, rederiving the covariance matrix estimate, and remeasuring the data vectors for each subset of the data. We find these tests cover a wide range of systematics in the catalog and recommend future surveys setup infrastructure to enable easy repetition of the processes in their end-to-end pipeline.
\end{itemize}

This work is part of a series of papers. \citetalias{paper1} details the shear catalog and all associated systematic checks while \citetalias{paper2} describes the redshift distribution and its associated validations/calibrations. The cosmological constraints from the \decade data, derived using the full analysis methodology discussed here, are presented in \citetalias{paper4}. Results from combining this \decade dataset with DES Y3, and with another $\approx 3,\!300\deg^2$ of \decade data in the southern Galactic cap, can be found in \href{\#cite.paper5}{Anbajagane \& Chang et al. (\citeyear{paper5})}.

Finally, we reiterate the last point from our summary: there is significant value in building analysis frameworks that can easily ingest object selection functions defined by a user, and then use it in all steps of the end-to-end cosmic shear pipeline. In this work, we have discussed such functionality in the context of systematics validation, where it is a vital tool in checking that different subsets of the catalog produce statistically consistent cosmology constraints. However, we note that such functionality inherently increases the power and applicability of these datasets to the broader community. For example, combinations of lensing data with other datasets may benefit from using different galaxy selection functions, or different definitions for the tomographic bins. Additional selections on the source galaxies can also improve the lensing-based mass estimate of galaxy clusters \citep[\eg][]{Rau:2024:PhotoZ}, while cross-correlations with other surveys/datasets often use only a subset of the photometric survey \citep[\eg][]{Shin:2019:Splashback, Omori:2023:CMBL, Chang2023, Sanchez:2023:tSZ, Anbajagane:2024:Shocks}, and this can be done more robustly by recalibrating the relevant subset of the data. Even in lensing-only analyses, we have found uses for alternative tomographic binning of the data \citep{Secco2022MassAp, Campos:2023:SOMPZ}, and also for using different selection functions to isolate data that may be uniquely robust to certain systematics \citep{Samuroff:2019:IA, McCullough:2024:BlueShear}. Building such functionality into our future pipelines will significantly improve the usability of lensing data, and enable the broader astrophysics and cosmology community to take better advantage of our precision datasets.

\section*{Acknowledgements}

DA is supported by the National Science Foundation (NSF) Graduate Research Fellowship under Grant No.\ DGE 1746045. 
CC is supported by the Henry Luce Foundation and Department of Energy (DOE) grant DE-SC0021949. 
The DECADE project is supported by NSF AST-2108168 and AST-2108169.
The DELVE Survey gratefully acknowledges support from Fermilab LDRD (L2019.011), the NASA {\it Fermi} Guest Investigator Program Cycle 9 (No.\ 91201), and the NSF (AST-2108168, AST-2108169, AST-2307126,  AST-2407526, AST-2407527, AST-2407528). This work was completed in part with resources provided by the University of Chicago’s Research Computing Center. The project that gave rise to these results received the support of a fellowship from "la Caixa" Foundation (ID 100010434). The fellowship code is LCF/BQ/PI23/11970028. C.E.M.-V. is supported by the international Gemini Observatory, a program of NSF NOIRLab, which is managed by the Association of Universities for Research in Astronomy (AURA) under a cooperative agreement with the U.S. National Science Foundation, on behalf of the Gemini partnership of Argentina, Brazil, Canada, Chile, the Republic of Korea, and the United States of America.

Funding for the DES Projects has been provided by the U.S. Department of Energy, the U.S. National Science Foundation, the Ministry of Science and Education of Spain, 
the Science and Technology Facilities Council of the United Kingdom, the Higher Education Funding Council for England, the National Center for Supercomputing 
Applications at the University of Illinois at Urbana-Champaign, the Kavli Institute of Cosmological Physics at the University of Chicago, 
the Center for Cosmology and Astro-Particle Physics at the Ohio State University,
the Mitchell Institute for Fundamental Physics and Astronomy at Texas A\&M University, Financiadora de Estudos e Projetos, 
Funda{\c c}{\~a}o Carlos Chagas Filho de Amparo {\`a} Pesquisa do Estado do Rio de Janeiro, Conselho Nacional de Desenvolvimento Cient{\'i}fico e Tecnol{\'o}gico and 
the Minist{\'e}rio da Ci{\^e}ncia, Tecnologia e Inova{\c c}{\~a}o, the Deutsche Forschungsgemeinschaft and the Collaborating Institutions in the Dark Energy Survey. 

The Collaborating Institutions are Argonne National Laboratory, the University of California at Santa Cruz, the University of Cambridge, Centro de Investigaciones Energ{\'e}ticas, 
Medioambientales y Tecnol{\'o}gicas-Madrid, the University of Chicago, University College London, the DES-Brazil Consortium, the University of Edinburgh, 
the Eidgen{\"o}ssische Technische Hochschule (ETH) Z{\"u}rich, 
Fermi National Accelerator Laboratory, the University of Illinois at Urbana-Champaign, the Institut de Ci{\`e}ncies de l'Espai (IEEC/CSIC), 
the Institut de F{\'i}sica d'Altes Energies, Lawrence Berkeley National Laboratory, the Ludwig-Maximilians Universit{\"a}t M{\"u}nchen and the associated Excellence Cluster Universe, 
the University of Michigan, NSF's NOIRLab, the University of Nottingham, The Ohio State University, the University of Pennsylvania, the University of Portsmouth, 
SLAC National Accelerator Laboratory, Stanford University, the University of Sussex, Texas A\&M University, and the OzDES Membership Consortium.

The DES data management system is supported by the National Science Foundation under Grant Numbers AST-1138766 and AST-1536171.
The DES participants from Spanish institutions are partially supported by MICINN under grants ESP2017-89838, PGC2018-094773, PGC2018-102021, SEV-2016-0588, SEV-2016-0597, and MDM-2015-0509, some of which include ERDF funds from the European Union. IFAE is partially funded by the CERCA program of the Generalitat de Catalunya.
Research leading to these results has received funding from the European Research
Council under the European Union's Seventh Framework Program (FP7/2007-2013) including ERC grant agreements 240672, 291329, and 306478.
We  acknowledge support from the Brazilian Instituto Nacional de Ci\^encia
e Tecnologia (INCT) do e-Universo (CNPq grant 465376/2014-2).

Based in part on observations at Cerro Tololo Inter-American Observatory at NSF's NOIRLab, which is managed by the Association of Universities for Research in Astronomy (AURA) under a cooperative agreement with the National Science Foundation.

This work has made use of data from the European Space Agency (ESA) mission {\it Gaia} (\url{https://www.cosmos.esa.int/gaia}), processed by the {\it Gaia} Data Processing and Analysis Consortium (DPAC, \url{https://www.cosmos.esa.int/web/gaia/dpac/consortium}).
Funding for the DPAC has been provided by national institutions, in particular the institutions participating in the {\it Gaia} Multilateral Agreement.

This paper is based on data collected at the Subaru Telescope and retrieved from the HSC data archive system, which is operated by the Subaru Telescope and Astronomy Data Center (ADC) at NAOJ. Data analysis was in part carried out with the cooperation of Center for Computational Astrophysics (CfCA), NAOJ. We are honored and grateful for the opportunity of observing the Universe from Maunakea, which has the cultural, historical and natural significance in Hawaii. 

This research uses services or data provided by the Astro Data Lab, which is part of the Community Science and Data Center (CSDC) Program of NSF NOIRLab. NOIRLab is operated by the Association of Universities for Research in Astronomy (AURA), Inc. under a cooperative agreement with the U.S. National Science Foundation.

This manuscript has been authored by Fermi Forward Discovery Group, LLC under Contract No.\ 89243024CSC000002 with the U.S. Department of Energy, Office of Science, Office of High Energy Physics.

All analysis in this work was enabled greatly by the following software: \textsc{Pandas} \citep{Mckinney2011pandas}, \textsc{NumPy} \citep{vanderWalt2011Numpy}, \textsc{SciPy} \citep{Virtanen2020Scipy}, \textsc{Matplotlib} \citep{Hunter2007Matplotlib}, and \textsc{Getdist} \citep{Lewis:2019:Getdist}. We have also used
the Astrophysics Data Service (\href{https://ui.adsabs.harvard.edu/}{ADS}) and \href{https://arxiv.org/}{\texttt{arXiv}} preprint repository extensively during this project and the writing of the paper.

\section*{Data Availability}

All catalogs and derived data products (data vectors, redshift distributions, calibrations etc.) for the cosmology analysis are now publicly available through the Noirlab Datalab portal \citep{Fitzpatrick:2014:DataLab, Nikutta:2020:DataLab} as well as through Globus and other avenues. Please visit \url{dhayaaanbajagane.github.io/data_release/decade} for a list of the available dataproducts and their corresponding data access. Our intention is to make all useful products immediately available to the community. Please reach out to DA if a data product of interest to you is not on the above list.

\bibliographystyle{mnras}
\bibliography{References}



\appendix
\section{Impact of PSF contamination and $B$-modes}\label{appx:Contam2Cosmo}

\begin{table}
    \begin{tabular}{c|ccc}
        \hline
        Bin & $\alpha$ & $\beta$ & $\eta$ \\
        \hline
        \hline
        1 & $0.0025 \pm 0.0070$ & $1.6347 \pm 0.1630$ & $-0.8548 \pm 2.2794$\\
        2 & $-0.0136 \pm 0.0077$ & $1.7490 \pm 0.1890$ & $-3.2707 \pm 2.0824$\\
        3 & $-0.0058 \pm 0.0100$ & $1.6712 \pm 0.1818$ & $-2.8773 \pm 2.9182$\\
        4 & $-0.0015 \pm 0.0107$ & $1.8128 \pm 0.2270$ & $0.8296 \pm 3.0527$\\
        \hline
    \end{tabular}
    \caption{The PSF coefficients per tomographic bin, as used in Equation \eqref{eqn:PSFcontam}, computed separately for each tomographic redshift bin using only galaxies in that bin. All measured coefficients are statistically consistent with expected values of $\alpha \lesssim 10^{-3}$ and $\beta, \eta \sim 1$. The method for computing the coefficients follow that describe in \citetalias{paper1} (see their Section 4.5).}
    \label{tab:abe_tomographic}
\end{table}

\begin{figure}
    \centering
    \includegraphics[width=1\columnwidth]{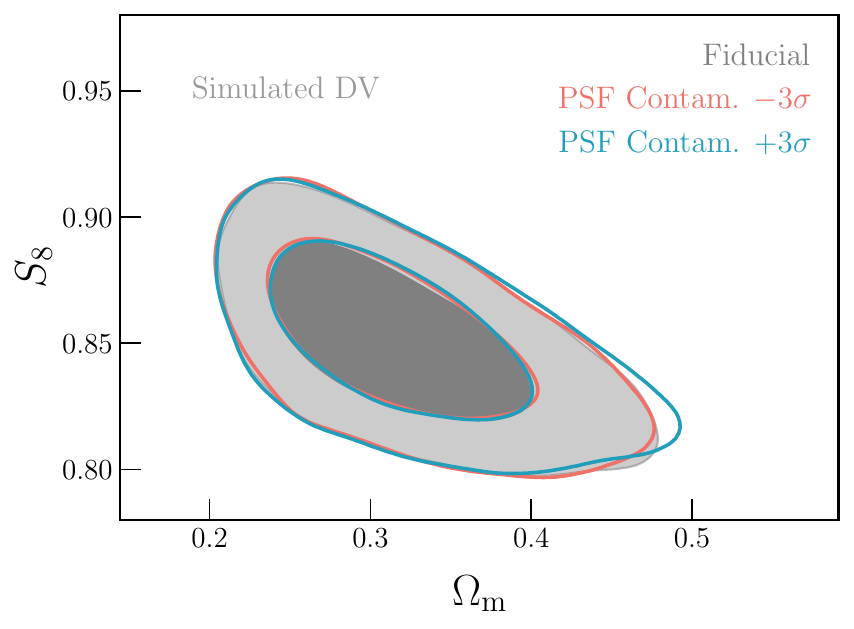}
    \caption{The impact of PSF contamination on the $S_8$-$\Omega_{\rm m}$ contours. We convert the distribution of $\alpha$, $\beta$, $\eta$ parameters (Equation \ref{eqn:PSFcontam}) into additive offsets to the data vector (DV), $\delta \xi_\pm^{\rm\, psf}$, and contaminate a simulated data vector with the $\pm 3\sigma$ values of the $\delta \xi_\pm^{\rm\, psf}$ distribution. The cosmology constraints are robust even after assuming extreme values of the PSF contamination.}
    \label{fig:psf_contam}
\end{figure}

\begin{figure}
    \centering
    \includegraphics[width=1\columnwidth]{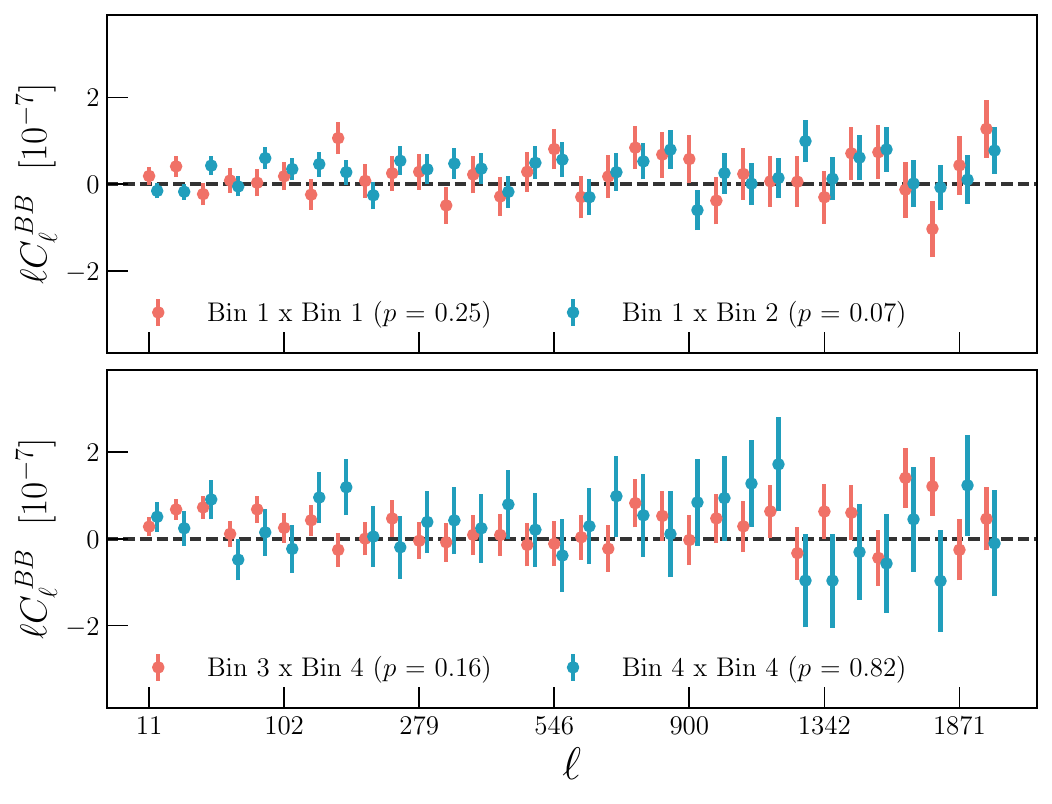}
    \caption{The angular power spectra of $B$-modes, $C_{\ell}^{BB}$, in the shape catalog, measured using galaxies from different tomographic bins (see legend). The measurements for all tomographic bin combinations are consistent with no $B$-mode signal. For brevity, we show only a subset of the measurements with the bin-combinations denoted in the legend.}
    \label{fig:bmode_tomographic}
\end{figure}

\subsection{PSF contamination}\label{appx:contam:PSF}

We denote the \textit{measured} two-component galaxy ellipticity in bin $a$, as $\boldsymbol{e}^{\rm obs}_a$. This ellipticity will have some contribution from untreated PSF contamination, $\boldsymbol{e}^{\rm obs}_a = \boldsymbol{e}^{\rm true}_a + \delta\boldsymbol{e}^{\rm psf}_a$ \citep[\eg][]{Paulin-Henriksson2008}. Now we can write the two-point correlation of the shapes as,
\begin{align}
    \langle \boldsymbol{e}^{\rm obs}_a \boldsymbol{e}^{\rm obs}_a \rangle & = \langle (\boldsymbol{e}^{\rm true}_a + \delta\boldsymbol{e}^{\rm psf}_a) (\boldsymbol{e}^{\rm true}_a + \delta\boldsymbol{e}^{\rm psf}_a) \rangle\nonumber\\
    & = \langle \boldsymbol{e}^{\rm true}_a\boldsymbol{e}^{\rm true}_a + 2\boldsymbol{e}^{\rm true}_a\delta\boldsymbol{e}^{\rm psf}_a + \delta\boldsymbol{e}^{\rm psf}_a\delta\boldsymbol{e}^{\rm psf}_a \rangle\nonumber\\
    & = \langle \boldsymbol{e}^{\rm true}_a\boldsymbol{e}^{\rm true}_a\rangle + \langle \delta\boldsymbol{e}^{\rm psf}_a\delta\boldsymbol{e}^{\rm psf}_a \rangle
\end{align}
where we follow previous shape measurement literature in assuming $\boldsymbol{e}^{\rm true}$ does not correlate with $\boldsymbol{e}^{\rm psf}$, \ie we assume $\langle \boldsymbol{e}^{\rm true} \boldsymbol{e}^{\rm psf} \rangle = 0$. He

Now, if we write the same for the cross-correlation,
\begin{align}
    \langle \boldsymbol{e}^{\rm obs}_a \boldsymbol{e}^{\rm obs}_b \rangle & = \langle (\boldsymbol{e}^{\rm true}_a + \delta\boldsymbol{e}^{\rm psf}_a) (\boldsymbol{e}^{\rm true}_b + \delta\boldsymbol{e}^{\rm psf}_b) \rangle\nonumber\\
    & = \langle \boldsymbol{e}^{\rm true}_a\boldsymbol{e}^{\rm true}_b + \boldsymbol{e}^{\rm true}_a\delta\boldsymbol{e}^{\rm psf}_b + \delta\boldsymbol{e}^{\rm psf}_a\boldsymbol{e}^{\rm true}_b + \delta\boldsymbol{e}^{\rm psf}_a\delta\boldsymbol{e}^{\rm psf}_b \rangle\nonumber\\
    & = \langle \boldsymbol{e}^{\rm true}_a\boldsymbol{e}^{\rm true}_b\rangle + \langle \delta\boldsymbol{e}^{\rm psf}_a\delta\boldsymbol{e}^{\rm psf}_b \rangle
\end{align}
for the same reasons mentioned above, $\boldsymbol{e}^{\rm true}$ does not correlate with $\boldsymbol{e}^{\rm psf}$ so the second and third term average to $0$ given $\langle\boldsymbol{e}^{\rm true}\rangle = 0$.

The PSF contribution to the measured ellipticity can be written as,
\begin{equation}
    \delta \boldsymbol{e}^{\rm psf}_a = \alpha_a \boldsymbol{e}^{\rm model} + \beta_a (\boldsymbol{e}^{\star} - \boldsymbol{e}^{\rm model}) + \eta_a \bigg(\boldsymbol{e}^{\star} \frac{T^\star - T^{\rm model}}{T^\star}\bigg)
\end{equation}
where $\boldsymbol{e}^{\star}$ and $T^\star$ are the ellipticity and size of the light profile, respectively, of a star, while $\boldsymbol{e}^{\rm model}$ and $T^{\rm model}$ are the same for the PSF model evaluated at the location of the star. The coefficients $\alpha_a, \beta_a, \eta_a$ are measured using only the galaxies in tomographic bin $a$. The quantities $\boldsymbol{e}^{\rm model}, \boldsymbol{e}^{\star}$, \textit{etc}. are not dependent on tomographic bin, but the coefficients associated with a given term \textit{does} depend on the bin. We can write the cross-term, $\langle \delta\boldsymbol{e}^{\rm psf}_a\delta\boldsymbol{e}^{\rm psf}_b \rangle$ as,
\begin{align}\label{eqn:PSFcontam}
\langle \delta \boldsymbol{e}^{\rm psf}_a  \delta \boldsymbol{e}^{\rm psf}_b \rangle
= \,\,& \alpha_a \alpha_b \rho_0 + \beta_a \beta_b \rho_1 + (\alpha_a\beta_b + \beta_a\alpha_b)\rho_2\nonumber\\ 
& + \eta_a\eta_b\rho_3 + (\eta_a\beta_b + \beta_a\eta_b)\rho_4  \nonumber\\
& + (\eta_a\alpha_b + \alpha_a\eta_b)\rho_5
\end{align}
where the $\rho_i$ are the Rowe statistics as described in \citetalias{paper1}.

We compute these bin-dependent coefficients using the same techniques as \citetalias{paper1} (see their Section 4.5) but now with only the source galaxies in tomographic bin $a$. The coefficient estimates are listed in Table \ref{tab:abe_tomographic}, and are consistent with the non-tomographic case presented in \citetalias{paper1}. We estimate the PSF contamination to the shear two-point data vector, $\delta \xi_\pm^{\rm\, psf}$, using Equation \eqref{eqn:PSFcontam} and the coefficients of Table \ref{tab:abe_tomographic}. For each bin, we generate a distribution of $\delta \xi_\pm^{\rm\, psf}$ contributions. This is done by sampling from the joint posterior of the $\alpha, \beta, \eta$ parameters in that bin, and generate a $\delta \xi_\pm^{\rm\, psf}$ per sample. We then compute the $99.7\%$ upper/lower bounds on $\delta \xi_\pm^{\rm\, psf}$ and use those two estimates to contaminate a simulated data vector, where the latter is generated using the same approach as Section \ref{sec:validation}. The constraints from the contaminated data vectors are shown in Figure \ref{fig:psf_contam}. The shifts in the cosmology are completely negligible, indicating that PSF contamination has negligible impact on the \decade cosmology constraints.

\subsection{$B$-modes}\label{appx:contam:Bmodes}

In \citetalias{paper1}, we verify that the \decade shape catalog does not contain any $B$-mode signal to within the statistical precision of the data. We now repeat this exercise, but perform the measurement for different tomographic bin combinations. Here, we have used the \textsc{Namaster} \citep{Alonso:2019:Namaster} package to measure the $B$-modes power spectra in our data. Our measurement follows the same approach as \citet{Doux:2022:ClsY3}, and we also generate measurements covariances using the same, simulation-based approach of that work.

Figure \ref{fig:bmode_tomographic} shows a subset of our results. We find no evidence of $B$-modes in any tomographic bin combination. All measurements not shown in Figure \ref{fig:bmode_tomographic} all have $p > 0.3$. Given we have no detection of $B$-mode contamination, we do not estimate a $\delta \xi_\pm^{\,B{\rm -mode}}$ contribution and propagate the impact of a $B$-mode contaminant to cosmology. This follows the choices in \citet{Amon2021}. Note that the harmonic-space estimator here differs from the real-space estimator used in \citetalias{paper1}. The former was chosen for this test as our original purpose was to contaminate the simulated data vector with $B$-modes (assuming a $B$-mode signal was detected) and this task requires as input the angular power spectra of $B$-modes. We have also verified that the non-tomographic measurement of the $B$-modes from \textsc{Namaster} is statistically consistent with no signal ($p = 0.41$) which itself is consistent with the result from the real-space estimator in \citetalias{paper1}.

\section{Additional results from catalog splits}\label{appx:SurveySplits}

\begin{table}
    \centering
    \begin{tabular}{c|cccc}
    \hline \rule{0pt}{10pt}
    Run & $a_2$ & $\eta_2$ & $\chi^2/N_{\rm dof}$ & $p$\\
    \hline \rule{0pt}{10pt}
    Airmass$^+$ & $-1.2^{+2.8}_{-2.5}$ & $2.1^{+1.9}_{-4.2}$ & 233.1/220 & 0.260 \\[3pt]
    Airmass$^-$ & $-2.0^{+4.7}_{-1.9}$ & $3.23^{+0.81}_{-2.2}$ & 268.1/220 & 0.015 \\[3pt]
    Seeing$^+$ & $-2.0^{+2.9}_{-2.1}$ & $2.6^{+1.5}_{-3.5}$ & 215.4/220 & 0.575 \\[3pt]
    Seeing$^-$ & $-3.0^{+1.2}_{-1.0}$ & $3.66^{+0.40}_{-0.64}$ & 287.0/220 & 0.002 \\[3pt]
    DCR-R.A.$^+$ & $-2.7^{+1.8}_{-1.4}$ & $3.25^{+0.78}_{-1.4}$ & 234.6/220 & 0.238 \\[3pt]
    DCR-R.A.$^-$ & $-0.6^{+1.8}_{-1.8}$ & $1.6^{+2.5}_{-4.4}$ & 252.5/220 & 0.066 \\[3pt]
    DCR-Dec.$^+$ & $-1.4^{+3.6}_{-2.2}$ & $2.6^{+1.5}_{-4.0}$ & 264.3/220 & 0.022 \\[3pt]
    DCR-Dec.$^-$ & $-2.3^{+5.3}_{-1.9}$ & $3.1^{+1.1}_{-2.1}$ & 236.4/220 & 0.214 \\[3pt]
    DCR$^+$-$\rm e_1$ & $-2.1^{+2.6}_{-2.0}$ & $2.9^{+1.1}_{-2.8}$ & 245.9/220 & 0.111 \\[3pt]
    DCR$^-$-$\rm e_1$ & $-2.2^{+1.6}_{-1.6}$ & $3.06^{+0.98}_{-2.1}$ & 233.2/220 & 0.259 \\[3pt]
    DCR$^+$-$\rm e_2$ & $-0.4^{+2.7}_{-2.3}$ & $2.0^{+2.0}_{-4.4}$ & 233.9/220 & 0.248 \\[3pt]
    DCR$^-$-$\rm e_2$ & $-2.4^{+2.1}_{-1.7}$ & $3.0^{+1.0}_{-2.4}$ & 212.6/220 & 0.627 \\[3pt]
    DECADE West & $-3.31^{+0.96}_{-0.72}$ & $3.48^{+0.54}_{-0.81}$ & 238.0/220 & 0.192 \\[3pt]
    DECADE East & $-0.4^{+1.8}_{-1.6}$ & $1.5^{+2.6}_{-5.2}$ & 256.1/220 & 0.048 \\[3pt]
    Pan-STARRS Calib. & $-2.9^{+1.1}_{-1.0}$ & $3.52^{+0.50}_{-0.78}$ & 256.8/220 & 0.045 \\[3pt]
    SkyMapper Calib. & $-0.1^{+3.2}_{-3.1}$ & $0.9^{+3.2}_{-4.4}$ & 234.3/220 & 0.242 \\[3pt]
    Skybright$^+$ & $-1.4^{+2.6}_{-2.3}$ & $2.1^{+1.9}_{-4.5}$ & 262.1/220 & 0.027 \\[3pt]
    Skybright$^-$ & $-1.2^{+2.0}_{-2.0}$ & $1.9^{+2.2}_{-3.8}$ & 251.7/220 & 0.070 \\[3pt]
    Skysigma$^+$ & $-2.5^{+6.5}_{-1.6}$ & $3.19^{+0.84}_{-1.3}$ & 265.3/220 & 0.020 \\[3pt]
    Skysigma$^-$ & $-1.3^{+2.0}_{-2.2}$ & $2.2^{+1.8}_{-4.4}$ & 262.5/220 & 0.026 \\[3pt]
    Exptime$^+$ & $-1.9^{+1.9}_{-1.9}$ & $3.2^{+1.0}_{-2.2}$ & 268.0/220 & 0.015 \\[3pt]
    Exptime$^-$ & $-1.7^{+1.9}_{-2.1}$ & $1.5^{+2.5}_{-3.3}$ & 204.6/220 & 0.764 \\[3pt]
    Maglim$^+$ & $-2.2^{+5.3}_{-1.9}$ & $3.39^{+0.75}_{-1.3}$ & 269.4/220 & 0.013 \\[3pt]
    Maglim$^-$ & $-2.7^{+1.7}_{-1.3}$ & $1.6^{+2.3}_{-2.7}$ & 220.1/220 & 0.486 \\[3pt]
    Nexp$^+$ & $-2.0^{+5.2}_{-1.9}$ & $3.45^{+0.68}_{-1.1}$ & 262.6/220 & 0.026 \\[3pt]
    Nexp$^-$ & $-2.4^{+1.9}_{-1.7}$ & $1.6^{+2.5}_{-3.2}$ & 221.3/220 & 0.462 \\[3pt]
    Extinction$^+$ & $-1.1^{+3.0}_{-2.2}$ & $2.2^{+1.9}_{-3.6}$ & 240.9/220 & 0.160 \\[3pt]
    Extinction$^-$ & $-1.8^{+1.8}_{-1.8}$ & $2.4^{+1.7}_{-3.2}$ & 252.2/220 & 0.067 \\[3pt]
    StarDensity$^+$ & $-1.2^{+3.4}_{-2.2}$ & $2.5^{+1.5}_{-3.9}$ & 230.4/220 & 0.302 \\[3pt]
    StarDensity$^-$ & $-2.4^{+1.7}_{-1.5}$ & $2.8^{+1.2}_{-2.5}$ & 239.6/220 & 0.173 \\[3pt]
    \hline
    \end{tabular}
    \caption{Constraints on two parameters from the TATT extensions to the IA model, the $\chi^2$ of the maximum posterior model, and the resulting p-value. Results are shown for all splits on survey property maps. The analogous results for the object property splits are in Table \ref{tab:IA_OPsplits}. Note that the posteriors are fairly non-Gaussian, and to enhance the interpretation of consistency/inconsistency of the results we utilize the $2\sigma$ (not $1\sigma$) bounds of the posterior. The p-values are not corrected for the ``look elsewhere'' effect. See text for details.}
    \label{tab:IA_SPsplits}
\end{table}

\begin{table}[]
    \centering
    \begin{tabular}{c|cccc}
    \hline \rule{0pt}{10pt}
    Run & $a_2$ & $\eta_2$ & $\chi^2/N_{\rm dof}$ & $p$\\
    \hline \rule{0pt}{10pt}
    $r - z < 0.85$ & $-1.2^{+4.2}_{-2.7}$ & $2.7^{+1.4}_{-3.5}$ & 253.7/220 & 0.059 \\[3pt]
    $r - z > 0.85$ & $-1.8^{+5.2}_{-2.1}$ & $2.0^{+2.1}_{-4.8}$ & 253.9/220 & 0.058 \\[3pt]
    $m_r < 22.5$ & $-1.3^{+5.2}_{-2.7}$ & $2.7^{+1.4}_{-2.6}$ & 271.9/220 & 0.010 \\[3pt]
    $m_r > 22.5$ & $-2.3^{+4.9}_{-1.8}$ & $2.4^{+1.7}_{-2.7}$ & 210.8/220 & 0.660 \\[3pt]
    $m_i < 22$ & $-0.9^{+3.2}_{-2.5}$ & $2.1^{+2.0}_{-3.8}$ & 215.1/220 & 0.581 \\[3pt]
    $m_i > 22$ & $-2.8^{+1.4}_{-1.2}$ & $2.2^{+1.8}_{-2.1}$ & 217.5/220 & 0.534 \\[3pt]
    $m_z < 21.6$ & $-0.3^{+2.5}_{-2.3}$ & $1.7^{+2.4}_{-4.5}$ & 217.2/220 & 0.540 \\[3pt]
    $m_z > 21.6$ & $-2.7^{+5.8}_{-1.4}$ & $3.11^{+0.93}_{-1.6}$ & 196.6/220 & 0.869 \\[3pt]
    SNR $< 27$ & $-1.8^{+4.6}_{-2.3}$ & $1.6^{+2.5}_{-3.9}$ & 195.1/220 & 0.886 \\[3pt]
    SNR $> 27$ & $-0.9^{+1.7}_{-1.8}$ & $1.9^{+2.1}_{-3.8}$ & 228.4/220 & 0.335 \\[3pt]
    T $< 0.48$ & $-0.6^{+2.9}_{-2.8}$ & $1.9^{+2.2}_{-4.7}$ & 202.2/220 & 0.800 \\[3pt]
    T $> 0.48$ & $-2.8^{+1.4}_{-1.2}$ & $3.0^{+1.0}_{-1.8}$ & 249.7/220 & 0.082 \\[3pt]
    Tpsf $< 0.45$ & $-2.4^{+6.2}_{-1.7}$ & $2.7^{+1.4}_{-2.2}$ & 235.5/220 & 0.226 \\[3pt]
    Tpsf $> 0.45$ & $-2.9^{+1.3}_{-1.1}$ & $3.65^{+0.42}_{-0.69}$ & 256.1/220 & 0.048 \\[3pt]
    T/Tpsf $< 1$ & $-1.1^{+4.1}_{-2.9}$ & $2.8^{+1.3}_{-3.8}$ & 242.9/220 & 0.139 \\[3pt]
    T/Tpsf $> 1$ & $-3.10^{+1.2}_{-0.93}$ & $2.9^{+1.1}_{-1.6}$ & 242.0/220 & 0.148 \\[3pt]
    \hline
    \end{tabular}
    \caption{Similar to Table \ref{tab:IA_SPsplits} but for splits performed using the object (not survey) properties. The p-values are not corrected for the ``look elsewhere'' effect. See text for details.}
    \label{tab:IA_OPsplits}
\end{table}

\begin{figure*}
    \centering
    \includegraphics[width=2\columnwidth]{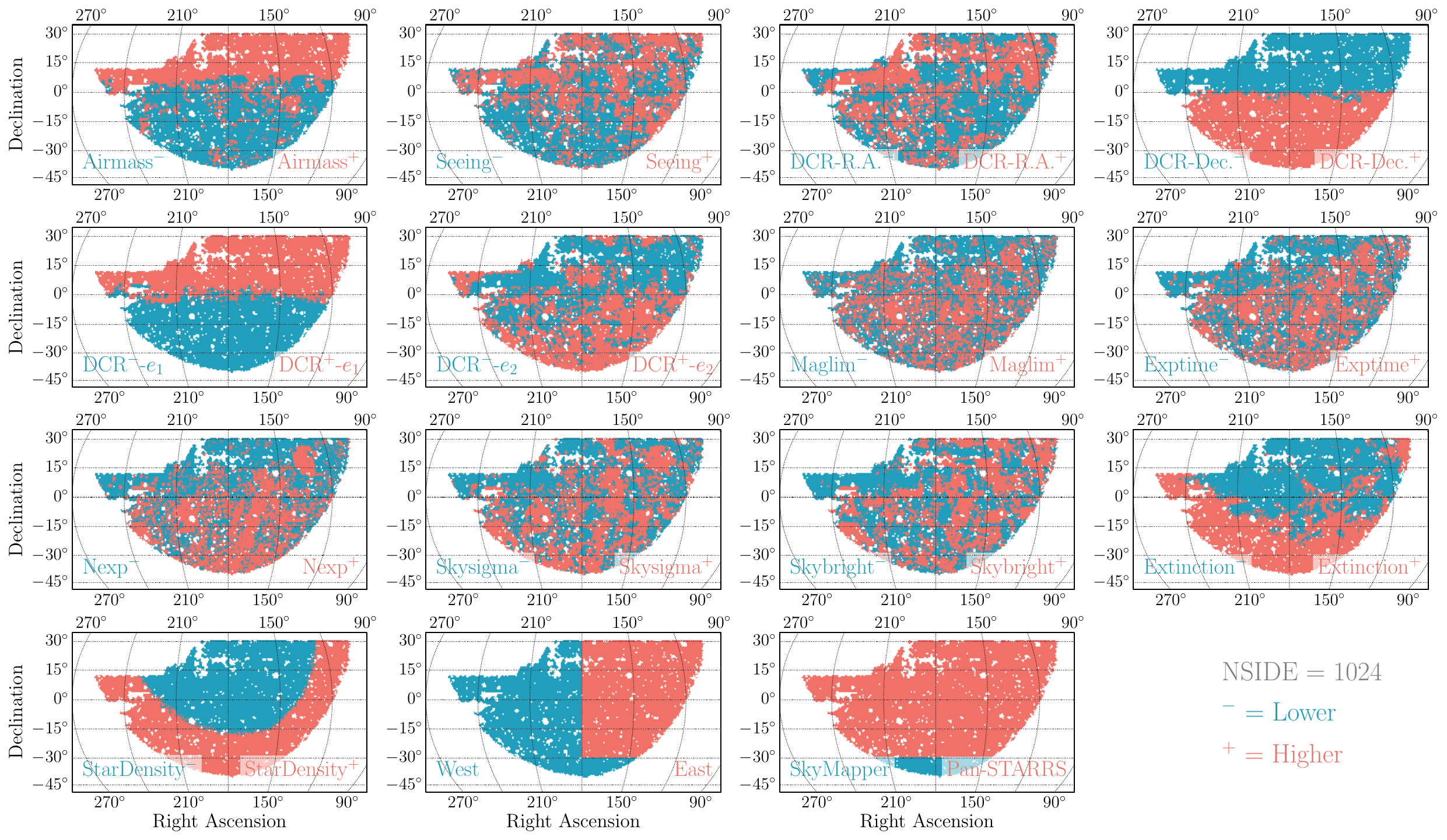}
    \caption{The area masks corresponding to each of the splits described in Section \ref{sec:SpatialInhomog:splits}. We use the same $\pm$ nomenclature to define regions where a given property is high/low. All survey property maps were generated at $\texttt{NSIDE} = 1024$. The western region (``West'') is not a strict split on longitude and includes some extra area that was part of the same, internal image processing campaign as the rest of the western patch.}
    \label{fig:SplitsMaps}
\end{figure*}

\begin{figure*}
    \centering
    \includegraphics[width=2\columnwidth]{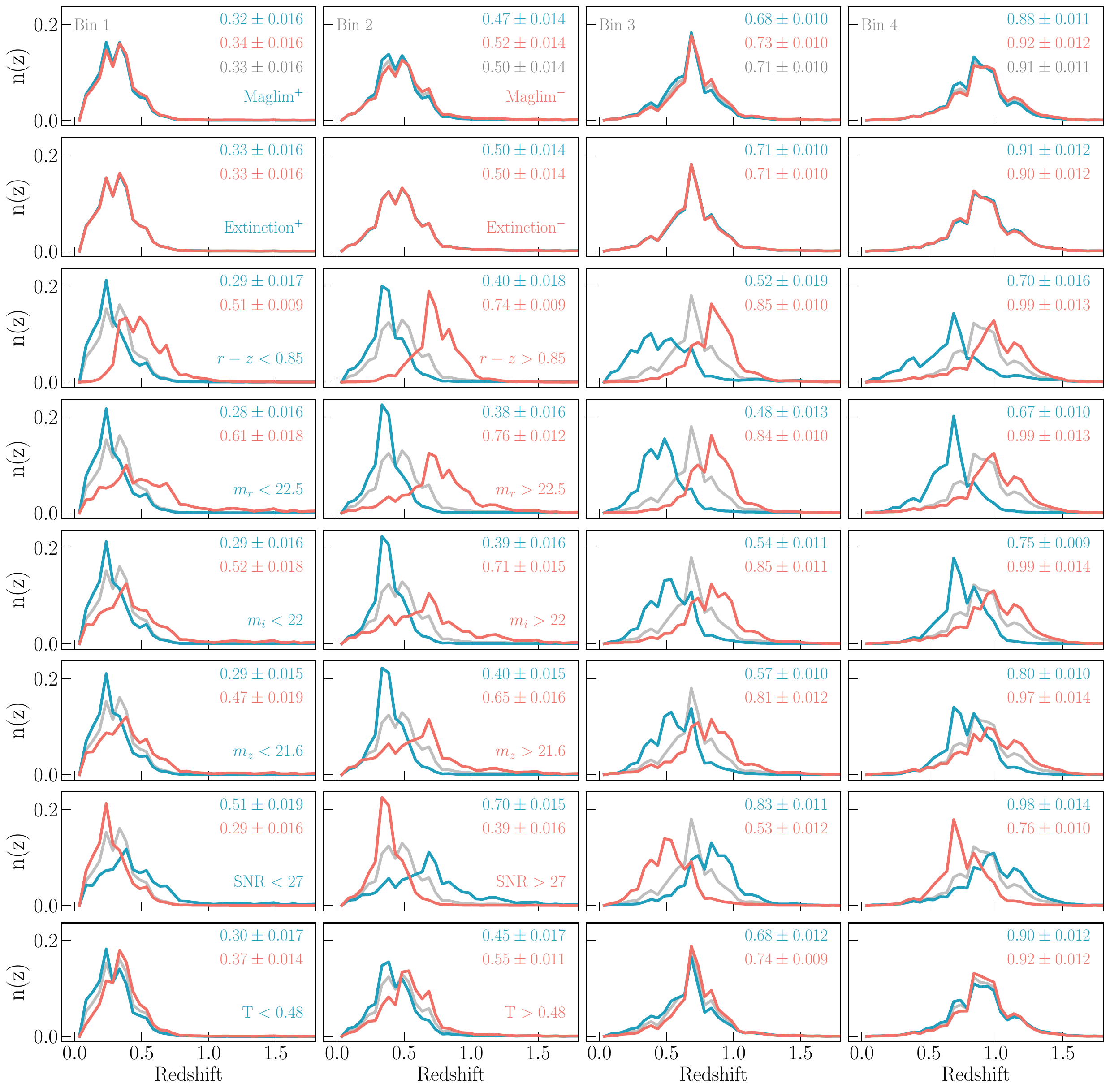}
    \caption{The mean redshift distributions, $n(z)$, for the four tomographic bins (columns) for some of the splits shown in Section \ref{sec:SpatialInhomog:splits} and \ref{sec:SpatialInhomog:Objsplits}. The fiducial redshift distribution is shown in gray. The text in the top shows the mean redshift, as well as the uncertainty on it as calibrated by the pipeline in \citetalias{paper2}. The text also denotes the split corresponding to each color. The statistics for the fiducial distributions are shown in top row. The area-based splits only lead to minimal changes in the mean $n(z)$ --- though they may have greater effects in causing spatially varying $n(z)$, which will not be highlighted in this figure which presents the area-averaged $n(z)$ --- while the object-based splits cause a clear separation in redshift. The mean redshifts for the fiducial case are, }
    \label{fig:SplitsRedshifts}
\end{figure*}

Section \ref{sec:SpatialInhomog:splits} and \ref{sec:SpatialInhomog:Objsplits} present cosmology constraints from different subsets of the footprint/catalog. Here, we denote some further details on the results in each split.

First, Figure \ref{fig:SplitsMaps} shows the survey footprint for each of the area-based splits that was analysed and presented earlier in Figure \ref{fig:SurveySplits}. The different colors distinguish the area spanned by a given subset and its complement. The splits on airmass are highly correlated with those on DCR, while there are similar correlations between the splits on exposure time, exposure count, and magnitude limit, and also between the two Milky Way properties (star density and extinction). All maps are shown at $\texttt{NSIDE} = 1024$, which is the resolution used to make the splits.

Second, Figure \ref{fig:SplitsRedshifts} shows the redshift distributions of the tomographic bins for a handful of different split definitions. The gray line in all panels show the distribution from the ``Fiducial'' setup as a reference. The area-based splits lead to some visible (but still minor) modification to the overall $n(z)$. The object-based splits --- which are done on object magnitudes or on quantities that correlate with object magnitudes --- lead to a more significant shift in the redshift distributions. The text in each panel lists the mean redshift of each distribution, and the uncertainty on the mean for that subset, estimated using the full uncertainty-quantification procedure detailed in \citetalias{paper2}.

Next, Table \ref{tab:IA_SPsplits} and Table \ref{tab:IA_OPsplits} list the constraints on two IA parameters --- $a_2$, the amplitude of the tidal torquing term, and $\eta_2$, the redshift dependence of this amplitude --- as well as the $\chi^2$ of the maximum posterior point in our chains alongside its associated p-value. The Seeing$^+$ result in Table \ref{tab:IA_SPsplits} is our worst-fit datavector, with $p = 0.002$. This barely passes our criteria of $p > 0.0015 (3\sigma)$ discussed in Appendix C of \citetalias{paper4}. However, we remain unconcerned as all estimates of the p-value do not include any corrections for the ``look elsewhere'' effect. In brief, we have split the same data into forty-six different subsets. The distribution of p-values will no longer follows a standard null distribution. The exact deviations can be quantified using a simulation-based approach, where one generates $N$ simulated datasets, measures datavectors for the forty-six subsets on it, and computes the resulting p-values. This calibration is unfeasible in our case as the simulations must have a high enough fidelity to reproduce all properties of the data (\eg the connection between airmass and object properties), and we must generate at least $N \sim 100$ realizations, which results in $4600$ tests/chains. Returning to the origin of this discussion --- the low p-value of the datavector best-fit for Seeing$^+$ --- we note that Figure 13 in \citetalias{paper1} finds no correlation between the mean shear and seeing (which is referred to in that Figure 13 as ``fwhm''). This indicates the seeing is not contaminating our measured shear.

Table \ref{tab:IA_SPsplits} and Table \ref{tab:IA_OPsplits} also lists the IA parameters of each result. This is motivated by our results in \citetalias{paper4}, where we find a non-zero amplitude for the tidal torquing term of the TATT IA model. We find here that the results of a given split and its complement are in good consistency with one another. The largest deviation is between the \decade East and West split, but is consistent within $2.4\sigma$. This is still completely consistent with just statistical fluctuations, and we reemphasize our point above about the ``look elsewhere'' effect, which can impact the parameter posteriors as well and generally boosts the measured significance. We also note that the $a_2$ constraints between the objects with $T / T_{\rm psf} \gtrless 1$ are consistent within $0.9\sigma$, which is a clean test of the impact of PSF systematics on this constraint as the amplitude of such systematics generally scales linearly with this ratio \citep{Paulin-Henriksson2008}. This measurement of minimal impact from PSF contamination is consistent with our results in Figure \ref{fig:psf_contam}, which show the minimal impact on derived cosmology constraints.

Finally, we address the impact of correlations on the consistency between constraints of a subset and of its complement. As mentioned before, the shape noise in the measurement from a subset and from its complement are completely uncorrelated as no source galaxy is shared between the two. However, the two subsets' galaxy samples can still cover similar (or the same) sky area and therefore have correlations due to cosmic variance. These correlations are generally subdominant relative to shape noise but we perform a quantitative test of this here. To do so, we generate 800 mock catalogs from the \textsc{CosmoGrid} simulation suite \citep{Kacprzak2023Cosmogrid}, using the same procedure as \citet[][see their Section 2.2]{Yamamoto2025}. We generate these catalogs for each subset, while accounting for the exact sky-area mask of the subset and for the $n(z)$ of all tomographic bins in that subset. This generates, for each simulation realization, a catalog for every subset of interest. The resulting set of catalogs can be used to study the correlation between the cosmology constraints of a subset and its complement.

We perform this test for two subsets: Maglim$^\pm$, where the sky-area mask is quite patchy (and therefore the two subsets can be more strongly correlated by cosmic variance), and the $m_r \gtrless 22.5$ split, where the catalogs have significantly different $n(z)$ but share the same sky area. As a result, we generate four variant datavectors per simulation realization. For each datavector (so, for each subset in each of the 800 realizations), we estimate the cosmology constraints using the maximum likelihood optimizer available in \textsc{Cosmosis}.\footnote{We use a maximum likelihood as our point estimate of $\Seight$ instead of generating a full bayesian posterior and taking the mean/median as the latter is computationally infeasible for this task. The former is a common approach for quoting constraints in lensing datasets \citep[\eg][]{Secco2021, Amon2021, DESKiDS2023}.} We then compute the correlation between the 800 estimates of $\Seight$ from a subset and from its complement. For Maglim$^\pm$, we see a correlation of $\rho = 0.18$ between the estimates and for $m_r \gtrless 22.5$, we see a correlation of $\rho = 0.28$. These are relatively mild correlations and only change the inferred parameter-shift significances in Section \ref{sec:SpatialInhomog} by 10-15\%. The correlations between a subset and its complement for other split definitions will be lower than those estimated above. The Maglim$^\pm$ is the patchiest split of all considered (see Figure \ref{fig:SplitsMaps}) and the $m_r \gtrless 22.5$ selection is highly correlated with the other selections on signal-to-noise, brightness, size \textit{etc}. used in Figure \ref{fig:SurveySplitsProperties}. In summary, the mild correlations between the constraints from a subset and its complement only cause minor ($\sim 10\%$) changes to the inferred significances of the parameter shifts between the two constraints, and do not change any of our qualitative discussion. Critically, all split tests continue to pass comfortably even if we account for this minor correlation.

\section{Validation of shape noise term in the covariance matrix model}\label{appx:cov}

\begin{figure*}
    \centering
    \includegraphics[width=2\columnwidth]{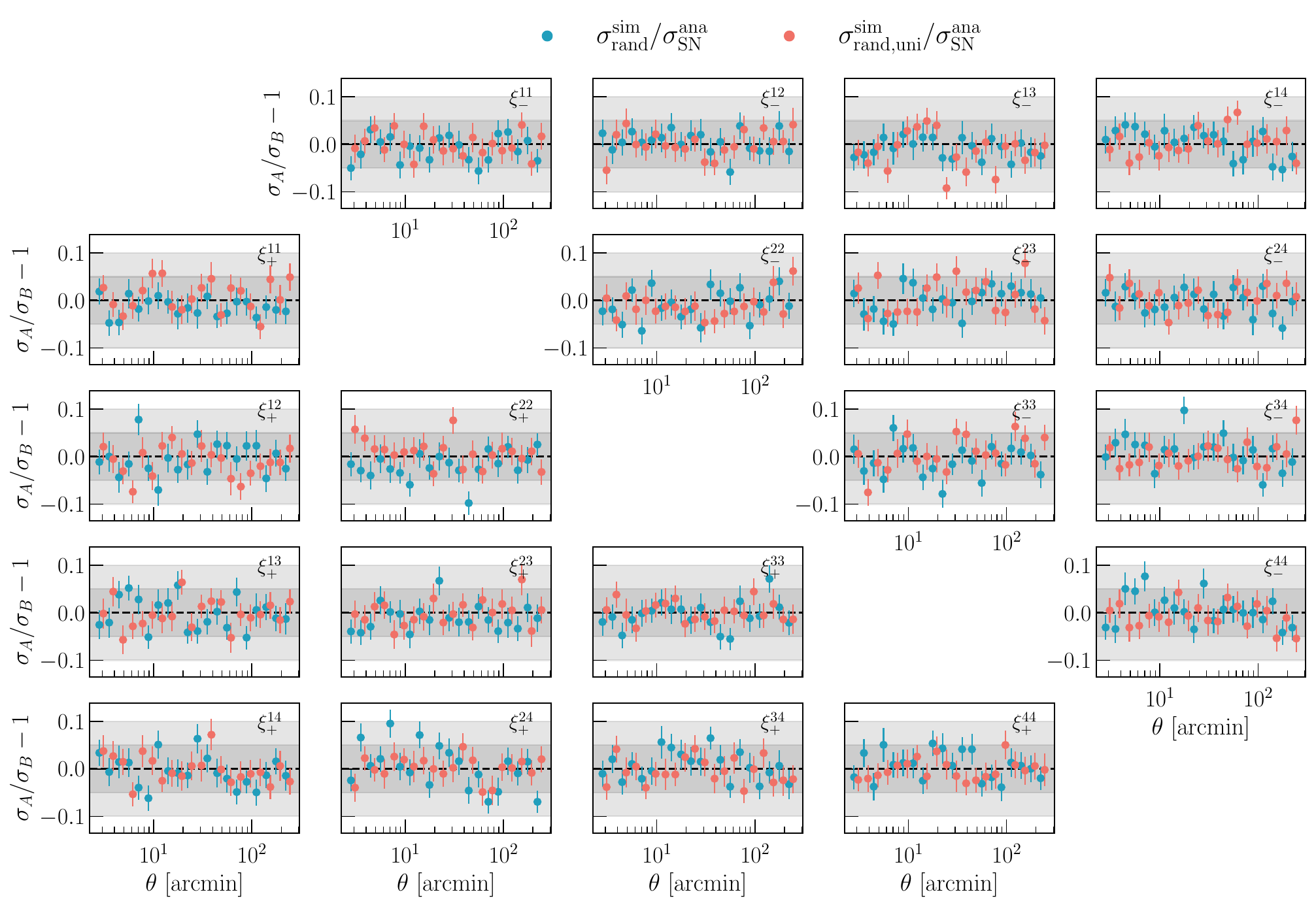}
    \caption{Ratios of the shape noise component in the covariance matrix between that measured from 600 simulations ($\sigma^{\rm sim}_{\rm SN}$) and analytical predictions ($\sigma^{\rm ana}_{\rm SN}$). We show the square root of the diagonal of the covariance matrix for each tomographic bin pair, and for two scenarios: (1) where we use the galaxy positions that are identical to the data and contain some level of spatial inhomogeneity (blue, $\sigma^{\rm sim}_{\rm rand}$) and (2) where we reposition each galaxy randomly inside the footprint to create a homogeneously distributed sample (red, $\sigma^{\rm sim}_{\rm rand, uni}$). Both sets of points scatter within 5\%, indicating that our analytical prescription of shape noise is accurate. We further find that the two sets of points do not show any qualitatively different behaviors, indicating that the spatial inhomogeneity in the galaxy number counts does not affect the shape noise component of the covariance significantly. Uncertainties on the simulation covariance are obtained from a leave-one-out jackknife across the 600 simulations.}
    \label{fig:CovTest_Shapenoise}
\end{figure*}

As mentioned in Section \ref{sec:cov}, the $\mathbf{C}_{\rm SN}$ term of the covariance matrix $\mathbf{C}$ includes the effect of the survey mask but not that of the variation in galaxy counts within said mask. However, we now show that the latter effect does not cause an appreciable shift in the covariance matrix estimates.

This validation is done by randomly rotating the measured ellipticities of the \decade shape catalog and computing the $\xi_\pm$ data vectors using \textsc{TreeCorr} \citep{Jarvis2004TreeCorr}. Such techniques have been used by \citet[][]{Troxel:2018:Cov} for validating the shape noise model. By repeating this rotation step for $\mathcal{O}(10^3)$ different realizations, we obtain a data-driven covariance matrix of $\xi_\pm$ that includes only shape noise, but accounts for the relevant spatial variations in the data by directly using the positions and shapes from the \decade catalog. We also perform a variant where the data vector is computed after removing the inhomogeneities, simply by reassigning \decade galaxies to new positions that now uniformly sample the survey mask. Figure \ref{fig:CovTest_Shapenoise} shows the ratio between the shape noise terms from the simulation-based method relative to those from the analytic method. This comparison is done for both the fiducial simulation-based covariance and the one with randomized galaxy positions; note that the analytic estimate is the same in either comparison. We find that in both cases, the analytic method matches the simulated one to within $5\%$. The fact that both sets of points are consistently within $5\%$ of the analytic estimate highlights that the spatial variation of the galaxy number density is a negligible factor in calculating the $\mathbf{C}_{\rm SN}$ term. The errorbars on the simulation covariance matrices are estimated using a leave-one-out jackknife method, \ie we remove one simulated data vector from the full set and generate the covariance matrix.

Finally, we have also compared the analytical covariance matrix (using all terms in Equation \ref{eqn:cov:contributions}) with a jackknife-based covariance matrix computed on the data. The main additional aspect that this checks --- beyond the test above --- is any significant discrepancy in the off-diagonal terms of the covariance matrix. However, the relative differences of the two estimates depend on the cosmological parameters assumed for the analytic estimate and therefore cannot be a rigorous test before unblinding our cosmology constraints. In addition, the jackknife method is known to be biased on large scales where the variation across jackknife regions is no longer independent \citep[\eg][]{Shirasaki:2017:JK}; we have also verified this limitation via simulated data vectors. As such, we performed this comparison only as an approximate check and do not show the results here given the limitations in their interpretability. 

\label{lastpage}
\end{document}